\documentclass[fleqn,usenatbib]{mnras}

\usepackage{newtxtext,newtxmath}

\usepackage[T1]{fontenc}
\usepackage{comment}

\DeclareRobustCommand{\VAN}[3]{#2}
\let\VANthebibliography\thebibliography
\def\thebibliography{\DeclareRobustCommand{\VAN}[3]{##3}\VANthebibliography}

\usepackage{graphicx}	
\usepackage{amsmath}	
\usepackage{gensymb}
\usepackage{physics}

\def\lsim{\lower.5ex\hbox{$\; \buildrel < \over \sim \;$}}
\def\gsim{\lower.5ex\hbox{$\; \buildrel > \over \sim \;$}}
\def\rg{r_{\rm g}}
\def\nel{n_{{\rm e}^-}}
\def\np{n_{\rm p}}
\def\mel{m_{{\rm e}}}
\def\mp{m_{\rm p}}
\def\sigmat{\sigma_{\rm T}}
\def\msol{\rm{M}_\odot}
\def\mdotsk{\dot{m}_{sk}}
\def\lpsd{l_{\rm{ps}}}
\def\cot{{\rm{cot}}}
\def\thsk{\theta_{{\rm{sk}}}}
\DeclareMathAlphabet\mathbfcal{OMS}{cmsy}{b}{n}
\def\xs{x_{\rm s}}
\def\xin{x_{\rm in}}
\def\xo{x_{\rm o}}
\def\hs{h_{\rm s}}
\def\er{E_{\rm rad}}
\def\rin{r_{\rm in}}
\def\vin{v_{\rm in}}
\def\thin{\Theta_{\rm in}}

\title[Radiatively driven jets]{Radiatively driven, time dependent bipolar outflows}

\author[]{
Raj Kishor Joshi$^{1, 2}$,
Indranil Chattopadhyay$^{1}$ \thanks{E-mail: indra@aries.res.in},
Lallan Yadav $^{2}$ 
\\
$^{1}$Aryabhatta Research Institute of Observational Sciences (ARIES), Manora Peak, Nainital 263001, India\\
$^{2}$Department of Physics, Deen Dayal Upadhyay Gorakhpur University, Gorakhpur\\
}

\date{Accepted XXX. Received YYY; in original form ZZZ}

\pubyear{2020}

\begin{document}
\label{firstpage}
\pagerange{\pageref{firstpage}--\pageref{lastpage}}
\maketitle

\begin{abstract}
We study the radiatively driven fluid jets around a non-rotating black hole. The radiation arising from the inner compact corona and outer sub-Keplerian part of the disc accelerates the jets. We obtain the steady state, semi-analytical, radiatively driven outflow solutions.
The thermodynamics of the outflow is described by a variable adiabatic index equation of state. We develop a TVD routine to investigate the time dependent behaviour of the radiatively driven bipolar outflow.
We inject with flow variables from the steady state outflow solutions
in the TVD code and allow the code to settle into steady state and match the numerical results with the steady state solution.
The radiation arising out of the accretion disc can provide a wide range of jet solutions, depending upon parameters like the intensity of disc, location of the inner corona etc. We induce the time dependence of the radiation field by inducing oscillation of the inner corona of the accretion disc. The radiation field then makes the bipolar outflow time dependent. We show that a non-steady radiation field arising out of disc oscillations can generate the internal shocks closer to the jet base.
Depending on the disc geometry, there might be transient shocks in the jet and there might be multiple non-stationary shocks in the jet, which
 are of much interest in jet physics.
\end{abstract}




\section{Introduction}

Astrophysical jets have been associated with wide range of objects such as active galactic nuclei (AGN), X-ray binaries (microquasars), and young stellar objects (YSOs). Stellar winds are the outward expansion of stellar atmosphere. However, black holes which reside at the center of microquasar or AGN/quasar does not have any atmosphere or a hard surface, hence the origin of jet must be accreting matter itself. Simultaneous X-ray and radio observations of microquasars have shown a strong correlation between the jets and spectral states of the accretion disc \citep{rsfp10,fgr10}, which also suggests that jets originate from the accretion disc but the entire disc may not participate in jet production as recent observations \citep{dl12} suggest that jets originate within a region of $100\,r_g$. The hot and ionized outflowing jet which originates from close vicinity of the compact object, ploughs through the intense radiation field of the accretion disc. The interaction of radiation field and plasma is not a new subject and equations of motion for radiation hydrodynamics were developed by many authors \citep{hs76,mm84,k98} and these equations have been used to study the radiatively driven winds and jets around compact objects. \citet{w74} showed that the radiation pressure arising out of high flux of soft X-rays can drive mass outflow from the outer regions of the accretion disc. \citet{icke80} studied the flow of particles above an alpha disc \citep{ss73} ignoring the radiation drag. Later in his seminal paper \citet{icke89} showed that the radiation drag ensures an upper limit on the terminal speed of plasma, which is around $0.45c$, termed as 'magic speed'. The Japanese group led by Jun Fukue has made a significant contribution to this field. Assuming a similar type of radiation field as considered by Icke, \citet{f96} studied the relativistic winds under the influence of radiation drag using the streamline approach (the dynamical equations are expressed by the streamline coordinates). However, the main problem was the collimation, as the winds gain angular momentum from the disc radiation field. To address the problem of collimation of jets, \citet{f99} studied the radiative jets confined by a disc corona. Later \cite{fth01} considered a hybrid disc with an inner advection dominated accretion flow (ADAF) \citep{nkh97} and an outer Keplerian disc (KD). \citet{ct95} considered a disc model by considering a mixture of matter with Keplerian and sub-Keplerian angular momentum and showed that the sub-Keplerian disc (SKD) can undergo a shock transition and due to extra heating in post shock region SKD and KD merge together to form a hot post shock disc (PSD). Numerical simulation of sub-Keplerian accretion disc \citep{mrc96,dcnm14,lcksr16} showed that the extra thermal gradient force present in the PSD, automatically generates the bipolar outflows.
The inner hot region or PSD may act as the illusive corona.
Whether the disc shock creates the inner, hot, torus like region
or some other mechanism do that is beyond the scope of this paper,
but suffice is to say that such hot torus like region has been proposed
by diverse researchers \citep{ct95,dove97,gzd97}.
\citeauthor{cc2000} (\citeyear{cc2000,cc00b,cc02b}) studied the interaction of intense radiation arising out of the PSD with outflowing jets  and showed that the jets can achieve the terminal speed in range $0.2c-0.3c$. Later \citet{c05} showed that the particle jets can be accelerated upto a terminal Lorentz factor $\gamma_T \geq 2$ and the radiation from two component disc \citep{ct95} provides significant collimation.   

In recent years there have been a large number of studies to investigate the propagation of relativistic jets and their interaction with ambient medium, effect of magnetic field on jets \citep{mm97,dh94, kbvk07, mrbfm10, wamkp14} but the numerical simulations of radiatively driven outflows are limited \citep{cc02a,csnr12, rvc21} probably because of the general consensus that radiation is not an efficient accelerating
agent \citep{ggmm02}. There are some simulations of line driven winds to study the effect of radiation on outflows \citep{psk2000,no17,ybl18} but the line forces are only effective when the temperature of wind is less than the ionization temperature ($T<10^5 K$). The recent investigations by \cite{vc17,vc18,vc19} have shown that the radiation can accelerate jets to relativistic terminal speeds and winds can also achieve mild to sub-relativistic speeds \citep{fa07,yf21,rvc21}, which suggests that the interaction of jet material with the radiation field becomes an important aspect to govern its dynamics. Also, the interaction of radiation comes into picture while explaining the internal shocks in the jets. 
In addition \citet{ftrt85,vc17,vc18,vc19} all showed that for inner disc geometry, standing shock can form close to the jet base.
The shocked region may accelerate the electrons and may produce non-thermal, high energy photons. It may be noted that, shocks are found to form in numerical simulation when a supersonic jet is launched into a denser
cold medium. However, these types of shocks are formed because of the interaction of a supersonic jet beam with ambient medium hence they form at a larger distance from the central object. 
In this paper, we would like to investigate whether the radiation can produce shocks when the jet travels through the intense radiation field as a fraction of radiation coming towards it can slow down the jet material and create the possibility of multiple sonic points and shocks in the flow \citep{ftrt85, vc17}. More interestingly, we would like to
see what happens of the accretion disc radiation field is itself time dependent. Due to the finite value of the speed of light, a change in the radiation field cannot be communicated to the entire length of the jet
at once. Hence the jet material farther out will not realise the radiation field has changed, although the inner part of the jet experiences a new radiation field. To study this we have written a new simulation code, using an equation of state of the gas with variable adiabatic index. We have computed the radiation field close to the axis of symmetry. We matched the simulation results with the analytical steady
state solution. The time dependent nature of the accretion disc, produces 
a time dependent radiation field, and thereby we study how radiation field may affect the jet. In particular, we want to investigate whether
a smooth jet can develop a shock due to the change in radiation field
of the accretion disc.

In section \ref{sec:equations} we present the governing equations and underlying assumptions. We also present a brief description CR EoS in section \ref{subsec:eos}. We present the methodology to obtain the solutions in section \ref{sec:method}. A brief description of the simulation code used is given in section \ref{subsec:code}. We present the time dependent as well as steady state solutions in section \ref{sec:results} and draw the concluding remarks in section \ref{sec:disc}.

\section{Assumptions and governing equations of jet}
\label{sec:equations}
We study the non-viscous and non-rotating outflows around a non-rotating black hole. The pseudo-Newtonian potential \citep{pw80} takes care of strong gravitational field around the black hole. The astrophysical jets remain collimated for long distances hence the transverse structure of jets is ignored and we have assumed a narrow conical geometry for the jet and all the flow variables are calculated on the axis. In this paper,
the jet-disc connection is not explored, instead we inject the jet into the computational domain with some temperature and radial velocity.
The accretion disc plays a supportive role by supplying the radiation which accelerates the jet. We describe the accretion disc in section \ref{sec:accretdisc}.
The equations of motion of radiation hydrodynamics have been investigated by many workers \citep{mm84,k98}. These equations, {for an optically thin jet}, correct up to first order in velocity are given as   

\begin{equation}
\frac{\partial \rho}{\partial t}+\mathbf{\nabla}.\left(\rho\mathbf{v}\right)=0\\
\label{eq:eqn_rhd1}
\end{equation}

\begin{equation}
\rho\frac{\partial \mathbf{v}}{\partial t}+\rho \left(\mathbf{v.\nabla}\right)\mathbf{v}=\mathbf{f}_g-\mathbf{\nabla} p-\rho_e\frac{\sigmat}{\mel c}\mathbf{R}\\
\label{eq:eqn_rhd2}
\end{equation}

\begin{equation}
\frac{\partial E}{\partial t}+\mathbf{\nabla}.\left[(E+p)\mathbf{v}\right]=\left(\rho_e v_i R_i+\rho v_i f_{g,i}\right)
\label{eq:eqn_rhd3}
\end{equation}
Where $\rho_e$ and $\rho$ are the leptonic mass density and total mass density of the flow, respectively. In equation \ref{eq:eqn_rhd3}, $E=\frac{1}{2}\rho v^2+e$ is the total energy density of the fluid, and $e$ is the thermal energy density. The gravitational force is represented by $\mathbf{f}_g$ and $\mathbf{R}$ is the net radiative contribution and the components of radiation term are given as   

\begin{equation}
R_i=F_i-v_i(\er+P_{ij})
\label{eq:rad_comp}
\end{equation}
$\er,\,F_i,\, {\rm{and}}\, P_{ij}$ represent the radiation energy density, components of radiation flux, and various components of radiation pressure tensor, respectively. The positive term
in equation (\ref{eq:rad_comp}) is radiative accelerating term, while
the negative term is the decelerating term. The decelerating term depends on $\er$ and $P_{ij}$, as well as $v_i$. It is for this reason the decelerating term is called 'radiation drag' term. Various moments of the radiation field is computed and is described in section \ref{sec:moments}. In this paper, we solve
equations (\ref{eq:eqn_rhd1}---\ref{eq:eqn_rhd3}). We obtain
the steady state semi-analytical solution of the jet similar to \citep{cc02b}, and then inject the flow variables at some injection point taken from those analytical solutions as inputs in the numerical simulation code.
\\

An additional equation, which relates $e,\,p,\,\rho$, known as equation of state (EoS) is required as closure relation in order to solve the set of equations \ref{eq:eqn_rhd1}, \ref{eq:eqn_rhd2}, and \ref{eq:eqn_rhd3}.

\subsection{Equation of state}
\label{subsec:eos}
As the jets travel through a long range and the temperature can vary over several orders in magnitudes and \cite{t48} showed that it is unphysical to consider a fixed $\Gamma$ EoS to describe the thermodynamics of these types of flows. We use an equation of state (EoS) for multispecies fluids, with variable adiabatic index ($\Gamma$) known as CR EoS \citep{cr09}.
\citet{vkmc15} showed that CR EoS approximates the exact EoS \citep{c39} very well. CR EoS has been used in variety of astrophysical problems \citep{jcry21, scl20, sc19, vc19, camrs14}. The EoS is given as 
\begin{equation}
e=\Sigma_i\left(n_im_ic^2+p_i\frac{9p_i+3n_im_ic^2}{3p_i+2n_im_ic^2}\right)
\label{eq:eos_old}
\end{equation}
The index i represents the different species of the fluid and c is the speed of light. In the unit system where $c=1$ equation \ref{eq:eos_old} can be represented in the form
\begin{equation}
 e=\rho f,
\label{eq:eos2}
\end{equation}
where,
\begin{equation}
f=1+(2-\xi)\Theta\left[\frac{9\Theta+6/\tau}{6\Theta+8/\tau}\right]+\xi\Theta\left[\frac{9\Theta+6/\eta\tau}{6\Theta+8/\eta\tau}\right]
\label{eq:eos3}
\end{equation}
In the above equations $\rho=\Sigma_i n_im_i=\nel \mel (2-\xi+\xi/\eta)$, where $\xi=\np/\nel$, $\eta=\mel/\mp$ and
$\nel$, $\np$, $\mel$ and $\mp$ are the electron number
density, the proton number density, the electron rest mass, and proton rest mass.
Moreover, $\Theta=p/\rho$ is a measure of temperature and $\tau=2-\xi+\xi/\eta$. 
The expression for specific enthalpy is given as 
\begin{equation}
h=(e+p)/\rho=f+\Theta;
\label{eq:enthalp}
\end{equation}
The polytropic index N is given as 
\begin{eqnarray}
N =  \rho \frac{\partial h}{\partial p}-1=\frac{\partial f}{\partial \Theta} 
 =6\left[(2-\xi)\frac{9\Theta^2+24\Theta/\tau+8/\tau^2}{(6\Theta+8/\tau)^2}\right] \nonumber\\ 
 +6\xi \left[\frac{9\Theta^2+24\Theta/(\eta \tau)+8/(\eta \tau)^2}{\{6\Theta+8/(\tau \eta)\}^2}\right]
\label{eq:poly}
\end{eqnarray}
We can easily infer from equation \ref{eq:poly} that polytropic index is  a function of $\Theta$ and $\xi$. It approaches asymptotic values
$N\rightarrow 3$ as $\Theta \gg 1$; while $N\rightarrow 3/2$ as $\Theta \ll 1$ at very high and low temperatures. And the adiabatic index $\Gamma$ is  
\begin{equation}
\Gamma=1+\frac{1}{N}
\label{eq:adiab}
\end{equation}

\subsection{Steady state equations of motion}
\label{sec:simulationcode}

In the steady state all the $\partial/\partial t$ terms vanish and the equations of motion admit analytical solutions.
The mass outflow rate, which can be obtained by integrating the continuity equation (\ref{eq:eqn_rhd1}) is given as  
\begin{equation}
{\dot M}_{out}=\rho v \mathcal{A}
\label{eq:cont}    
\end{equation}
For a narrow conical jet the cross section of jet $\mathcal{A}\propto r^2$.\\
The first law of thermodynamics is given as 
\begin{equation}
\frac{d}{dr}\left(\frac{e}{\rho}\right)=\frac{p}{\rho^2}\frac{dp}{dr}
\label{eq:energy_eqn}    
\end{equation}

As the energy equation \ref{eq:energy_eqn} does not have any source or sink term, flow will be isentropic, and equation \ref{eq:energy_eqn} can be integrated along with equation \ref{eq:eos2} to obtain isentropic relation given as 
\begin{equation}
\rho=\mathcal{C}\Theta^{3/2}\left(3\Theta+4/\tau\right)^{k_1}\left(3\Theta+4/\eta\tau\right)^{k_2}\rm{exp(k_3)}
\label{eq:isentropic}    
\end{equation}
Where\begin{equation}
k_1=\frac{3}{4(2-\xi)},\, k_2=\frac{3\xi}{4},\, k3=-\frac{3}{\tau}\left[\frac{2-\xi}{3\Theta+4/\tau}+\frac{\xi}{\eta(3\Theta+4/\eta\tau)}\right]    
\end{equation} 
We can use equation \ref{eq:isentropic} and equation \ref{eq:cont} to obtain the expression for entropy accretion rate which is a constant of motion as
\begin{equation}
\dot{\mathcal{M}}={\rm{exp}}(k_3)\Theta^{3/2}\left(3\Theta+4/\tau\right)^{k_1}\left(3\Theta+4/\eta\tau\right)^{k_2}v\,z^2
\label{eq:scriptmdot}    
\end{equation}
We can use the first law of thermodynamics (energy conservation equation) \ref{eq:energy_eqn} and the EoS \ref{eq:eos2} to obtain the temperature gradient in jet as
\begin{equation}
\frac{d\Theta}{dr}=-\frac{\Theta}{N}\left(\frac{2}{r}+\frac{1}{v}\frac{dv}{dr}\right)
\label{eq:dthetadr}    
\end{equation}
And momentum balance equation \ref{eq:eqn_rhd2} with the help of equations \ref{eq:eos2}, \ref{eq:dthetadr} can be expressed as
\begin{equation}
\frac{dv}{dr}=\frac{2a^2/r-0.5/(r-1)^2+\mathcal{F}-v(\mathcal{E+P})}{v-a^2/v}
\label{eq:dvdr}
\end{equation}
Where $a$ is adiabatic sound speed given as 
\begin{equation}
a^2=\Gamma\Theta    
\label{eq:sspeed}
\end{equation}
In addition, $\mathcal{F}=\sigma_T F_r/(\mel c)$, $\mathcal{E}=
\sigma_T \er/( \mel c)$ and $\mathcal{P}=\sigma_T P_{rr}/( \mel c)$.
To obtain the solution we need to simultaneously integrate equations \ref{eq:dthetadr} and \ref{eq:dvdr}. As the jet material in the vicinity of the compact object is hot and slower, or in other words, jet is subsonic at the base, while far away from the central object it becomes less hot and the radiative energy drives the jet to high speed and jet becomes supersonic. Which means that at some point jet would become transonic and that point is known as the sonic point ($r_c$). At the sonic point flow velocity becomes equal to the sound speed and the velocity gradient $dv/dr\rightarrow 0$. The sound speed at the sonic point is given as
\begin{equation}
a_c=\frac{0.5r_c(\mathcal{E}_{c}+\mathcal{P}_{c})+\sqrt{(0.5r_c(\mathcal{E}_{c}+\mathcal{P}_{c}))^2-2r_c\mathcal{F}+r_c/(r_c-1)^2}}{2}
\label{eq:sp_condn}
\end{equation}
To calculate $dv/dr|_c$ we use the L' Hospital's rule and solve the resulting quadratic equation for $dv/dr|_c$. The quadratic equation can have two complex roots or two real roots. The real roots but with opposite signs produce X type sonic points and solutions passing through these sonic points are physical. We start the numerical integration of equations \ref{eq:dthetadr} and \ref{eq:dvdr} from sonic points and integrate outward and inward to obtain the full solution. 

\subsection{Simulation code and numerical method}
\label{subsec:code}

Our simulation code is based on the total variation diminishing (TVD) scheme introduced by \cite{h83}. TVD scheme is an Eulerian, second-order accurate, finite difference scheme. Simulation codes based on TVD schemes are robust and efficiently capture shocks and these codes have been  extensively used to study wide range of astrophysical problems \citep{rokc83, lrc11, csnr12, rvc21} The spatial and temporal evolution of conserved quantities $\rho,\,\rho v\, \mbox{and}\, E$ is computed using Roe type Riemann solver. Second order accuracy is obtained by first modifying the flux function and then applying a non-oscillatory
first-order accurate scheme \citep[see][for details]{h83,rokc83}. The equations of motion (equations \ref{eq:eqn_rhd1}---\ref{eq:eqn_rhd3}) in the conservative form for {one dimensional spherical symmetric flow are given as}  
\begin{equation}
\frac{\partial \vb*{q} }{\partial t}+\frac{1}{r^2}\frac{\partial (r^2 \vb*{F}_1)}{\partial r}+\frac{\partial \vb*{F}_2 }{\partial r}=\vb*{S}
\label{eq:rhd_conservative}
\end{equation}
Where $\vb*{q}$ is the state vector given as 

\begin{equation}
\vb*{q}=\begin{pmatrix}
		\rho \\
		\rho v \\
		E	
		\end{pmatrix}
\label{eq:state_vect}
\end{equation}
And the fluxes are given as 

\begin{equation}
\vb*{F}_1=\begin{pmatrix} \rho v \\ \rho v^2 \\ (E+p)v \end{pmatrix} \,,\, \vb*{F}_2=\begin{pmatrix} 0\\p\\0 \end{pmatrix}
\label{eq:fluxes}
\end{equation}
 
The source function $\vb*{S}$ is given as 

\begin{equation}
\vb*{S}=\begin{pmatrix}
		0 \\
		f_{g,r} -\rho_e \mathcal{R}_r \\
		f_{g,r} v  - \rho_e v \mathcal{R}_r	
		\end{pmatrix}
\label{eq:source}
\end{equation}

The gravitational force $f_{g,r}$ acting on radial direction is given as 

\begin{equation}
f_{g,r}=\frac{\rho}{2(r-1)^2}
\label{eq:gravity_radial}
\end{equation}
The contribution of radiative terms $\mathcal{R}_r$ from equation
\ref{eq:rad_comp} is 
\begin{equation}
\mathcal{R}_r=\mathcal{F}-v(\mathcal{E}+\mathcal{P})
\label{eq:radn_terms_2} 
\end{equation}
The gravity and radiative moments are updated as source terms in the simulation code. 

The simulation code is the usual TVD routine, with eigenvalues given by
\begin{equation}
\lambda_{1,3}=v \pm a;~~ \lambda_{2,}=v
\end{equation}
The form of right ($\tilde{\rm \bf R}_{1,2,3}$), left
($\tilde{\rm \bf L}_{1,2,3}$) eigen vectors, computation of the fluxes are exactly same as usual TVD routines \citep[for details see][]{h83,rokc83}. The updating the state vector ${\bf q}^n$ to ${\bf q}^{n+1}$ follows the procedure of \citet{h83,rokc83}
\begin{equation}
 {\bf L_x}{\bf q}^n=\mathbf{q}^n-\frac{\Delta t^n}{\Delta x}\left({\bar f}_{x,i+1/2}-{\bar f}_{x,i-1/2}\right),
 \label{eq:qupdate}
\end{equation}
\begin{equation}
{\bar f}_{x,i+1/2}=\frac{1}{2}[\mathbf{F}(\mathbf{q}_i^n)+\mathbf{F}
(\mathbf{q}_{i+1}^n)]-\frac{\Delta x}{2\Delta t^n}\Sigma_{k=1}^3
\beta_{k, i+1/2}\mathbf{\tilde R}^n_{k,i+1/2},
\end{equation}
\begin{equation}
\beta_{k, i+1/2}=Q_k\left(\frac{\Delta t^n}{\Delta x} \lambda^n_{k,i+1/2}+\gamma_{k,i+1/2}\right)\alpha_{k,i+1/2}-(g_{k,i}+g_{k,i+1}),
\end{equation}
\begin{equation}
\alpha_{k,i+1/2}=\mathbf{\tilde L}^n_{k,i+1/2}.(\mathbf{q}^n_{i+1}-
\mathbf{q}^n_{i-1}),
\end{equation}

\begin{equation}
\gamma_{k,i+1/2}= \left\{
\begin{array}{lcl}
\left(g_{k,i+1}-g_{k,i}\right)/\alpha_{k,i+1/2} & \mathrm{for} &
\alpha_{k,i+1/2}\neq0, \\
0 & \mathrm{for} & \alpha_{k,i+1/2}=0,
\end{array}
\right.
\end{equation}
\begin{eqnarray}
 g_{k,i} = \mathrm{sign}(\tilde{g}_{k,i+1/2})\mathrm{max}
\{0,\mathrm{min}[|\tilde{g}_{k,i+1/2}|, \\ \nonumber
\mathrm{sign}(\tilde{g}_{k,i+1/2})\tilde{g}_{k,i-1/2}]\},
\end{eqnarray}
\begin{equation}
\tilde{g}_{k,i+1/2} = \frac{1}{2}\left[Q_k(\frac{\Delta t^n}{\Delta x}
a_{k,i+1/2}^n)-\left(\frac{\Delta t^n}{\Delta x}a_{k,i+1/2}^n\right)^2
\right]\alpha_{k,i+1/2},
\end{equation}
\begin{equation}
 Q_k(x) = \left\{
\begin{array}{lcl}
x^2/4\varepsilon_k+\varepsilon_k & \mathrm{for} & |x|<2\varepsilon_k, \\
|x| & \mathrm{for} & |x|\geq2\varepsilon_k.
\end{array}
\right.
\end{equation}

Updated values of $\mathbf{q} \equiv (\rho,~\rho v,~\&~E)$ obtained by
solving equation \ref{eq:qupdate}, describe how the jet fluid advances in time from one point in space to the other. 
However, since
the EoS (equation \ref{eq:eos2}) is not a linear function of the pressure $p$, therefore, even after updating the state vectors ${\bf q}$, one needs to solve a cubic equation to obtain the value of $\Theta$ and from there the value of $p$ at each cell centre at every time.
Recalling the definition of $E=\rho v^2/2+e$ and equation \ref{eq:eos2}, we can write
\begin{equation}
 f =\frac{E}{\rho}-\frac{\rho^2 v^2}{2 \rho^2}=\mathcal{C}=\mbox{known
 from equation \ref{eq:qupdate}}
 \label{eq:retrivp1}
\end{equation}
Combining equation \ref{eq:eos3} with equation \ref{eq:retrivp1}, we obtain the cubic equation of $\Theta$
\begin{equation}
27\Theta^3+9K_1 \Theta^2+12K_2 \Theta +\frac{16(1-\mathcal{C})}{\eta \tau^2}=0,
\label{eq:retrivp2}
\end{equation}
where, 
\begin{eqnarray*}
K_1=\frac{(1+2/\eta)(2-\xi)+\xi(1/\eta+2)}{\tau}+(1-\mathcal{C}) \\
K_2=\frac{2}{\eta \tau^2}+\frac{(1-\mathcal{C})}{\tau}\left(1+
\frac{1}{\eta}\right)
\end{eqnarray*}
Equation \ref{eq:retrivp2} admits analytical solution.
Following the standard method \citep{as70} to solve cubic equation, we found that
the CR EoS admits only one unique root.
Defining $b_1=9K_1/27;~~b_2=12K2/27;~\&~ b_3=16(1-\mathcal{C})/(27 \eta \tau^2)$,
then
\begin{equation}
 Q=(3b_2-b^2)/9;~~\&~~ T=(9b_1b_2--27b_3-2B_1^3)/54
\end{equation}
and a discriminant
\begin{equation}
\Xi= Q^3+T^2
\end{equation}
The number of roots of equation \ref{eq:retrivp2} will depend on the value of $\Xi$. The CR EoS results $\Xi <0,~\&~ Q<0$, which means three real
but unequal root. The root of equation \ref{eq:retrivp2} is
\begin{equation}
\Theta=2\sqrt{-Q}{\rm cos}{\frac{\tilde \theta}{3}}-\frac{b_1}{3};\mbox{ where, } {\rm cos}{\tilde \theta}=\frac{T}{\sqrt{-Q^3}}
\label{eq:thetrut}
\end{equation}

Once we know $\Theta$ (equation \ref{eq:thetrut}), we retrieve the pressure from
the definition of $\Theta$ i.e., $p=\rho \Theta$.

\subsection{Geometry of accretion disc}
\label{sec:accretdisc}

\begin{figure}
\includegraphics[width=\columnwidth]{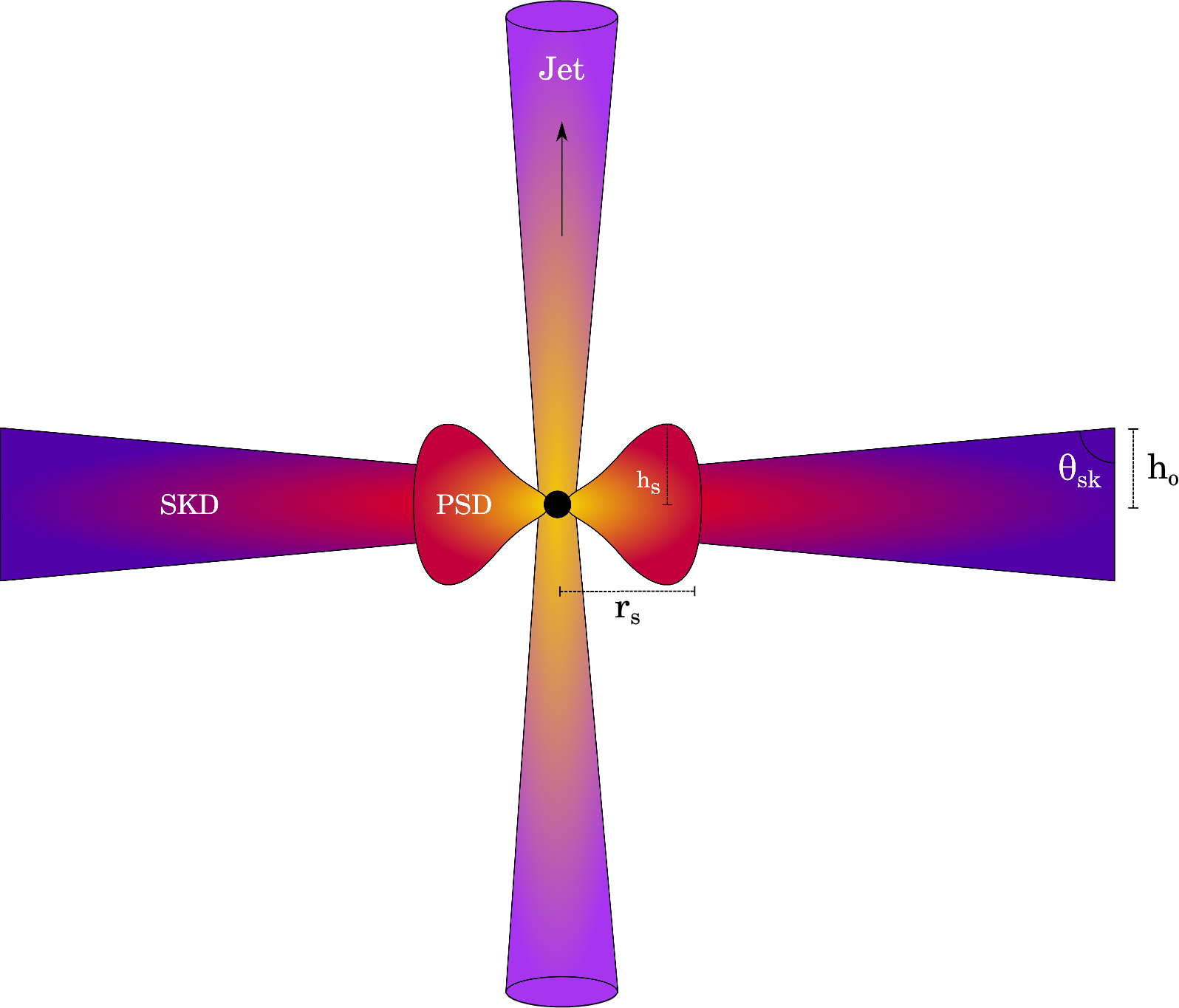}
\caption{Schematic representation of disc-jet system.}
\label{fig:disc_cartun}
\end{figure}

As we have mentioned before the accretion disc is the source of radiation only and is not part of the computational domain. We have considered the advective accretion disc \citep{f87, c89, ck16}. Figure \ref{fig:disc_cartun} shows various components of the disc. This disc structure is supposed to mimic the hard to hard-intermediate spectral states. The Keplerian disc (KD) is flanked by sub-Keplerian disc (SKD). SKD and KD merge at $\xs$ to form a single component, geometrically thick post shock disc (PSD) which is the source of hard photons. {Advective disc may be optically thin or internediate depending on the accretion rate. Similarly, advective discs depending on the accretion rate might be radiatively inefficient or moderately efficient \citep{scl20}. One may therefore call the advective discs are optically `slim' \citep{cdc04}. In particular, the cumulative optical  depth in the vertical direction considered in this paper is $\lsim 1$.} In principle, the inner edge of PSD $(\xin)$ should be at the horizon, but we have taken it to be at $1.5r_g$ for calculating the radiative moments because the region very close to the horizon is expected to emit little amount of radiation. The shock location $\xs$ is the inner edge of SKD. A large number of numerical simulations have shown that SKD is flatter than PSD \citep{gc13,mdc94}, so we have taken a semi-vertical angle of $85^{\rm o}$ for SKD. The outer edge of the disc is taken to be at $\xo=3500r_g$ and the intercept of SKD surface on the jet axis is $d_0=0.4\hs$, where $\hs$ is the shock height. Numerical simulations \citep{dcnm14, lcksr16} show that the height to radius ratio for corona or PSD can vary between 1.5-10. We have assumed that SKD emits via synchrotron process and the velocity and temperature profiles are required to calculate the SKD intensity. The details of obtaining velocity and temperature profile are presented in \cite{vkmc15}.

The temperature at $\xo$ is taken to be $\Theta_{\rm o}=0.2$ and injection velocity is assumed as $0.001$. The angular momentum $\lambda$ of the disc is 1.7. The PSD itself emits photons via synchrotron and bremsstrahlung processes but also inverse-Comptonize these photons and photons intercepted from SKD and KD. The KD emits thermal photons \citep{ss73}. It has been shown that the radiative moments from KD are very weak in comparison to PSD and SKD moments \citep{c05,vkmc15}, hence we have ignored the contribution from KD in this paper. 
\begin{figure}
\includegraphics[width=\columnwidth]{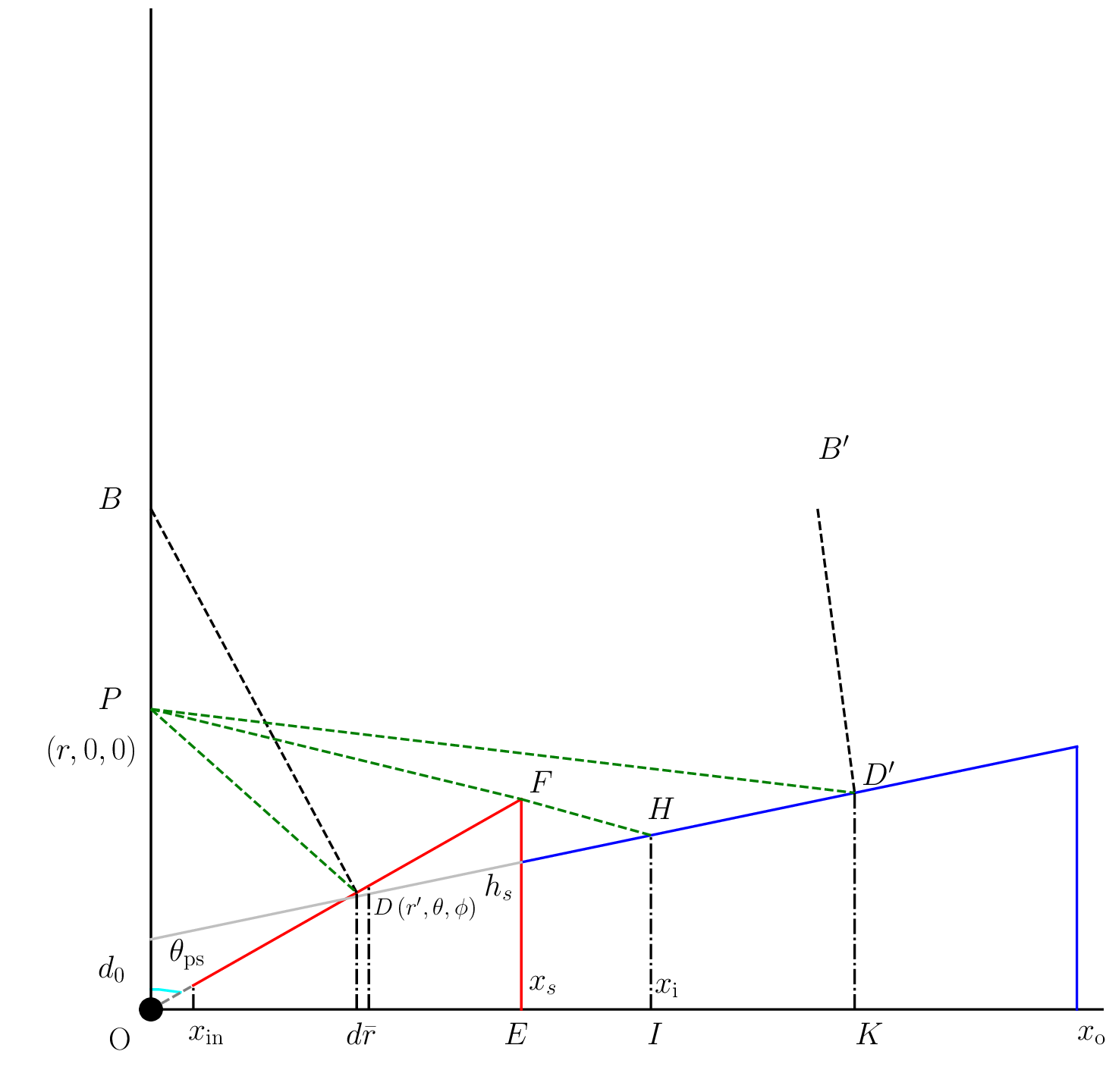}
\caption{Cartoon of the cross-section of the disc. The red portion is the PSD, which is plotted upto $\xin=1.5\rg$. The outer edge of PSD is shock location OE$=\xs$ and FE$=\hs$. P is the field point where radiative moments are computed. D is the source point on PSD and D$^\prime$ is the source point on SKD. BD is orthogonal at D, while B$^\prime$D$^\prime$ is orthogonal at D$^\prime$. Inner edge of SKD is $\xs$, but due to the shadow effect of PSD, the point P will only see OI$=x_i$ as the inner edge
of SKD. Outer edge of the disc is $\xo$.  }
\label{fig:compo}
\end{figure}

\subsection{Radiative moments above SKD and PSD}
\label{sec:moments}
To obtain the radiative moments we need the radiative intensities of different disc components (Fig. \ref{fig:disc_cartun}).
We assume that the synchrotron emission is the dominant mechanism in SKD. We also assume that there is stochastic magnetic field in SKD and the ratio of magnetic pressure and gas pressure is assumed to be constant

\begin{equation}
\beta=\frac{p_{\rm mag}}{p_{\rm gas}}=\frac{B^2_{\rm sk}/8\pi}{n_{\rm sk}kT_{\rm sk}}    
\end{equation}
Where $n_{\rm sk}$ and $T_{\rm sk}$ are local number density and temperature respectively.\\
The SKD intensity is given by \citep{st83, vkmc15}
\begin{equation}
I_{\rm sk}=\left[\frac{16}{3}\frac{e^2}{c}\left(\frac{eB_{\rm sk}}{\mel c}\right)^2\Theta_{\rm sk}^2n_{\rm sk}\right]\frac{\left(d_0\,{\rm sin\theta_{sk}}+x\,{\rm cos\theta_{sk}}\right)}{3}\,{\rm erg\, cm^{-2} s^{-1}}    
\end{equation}

Where $B_{\rm sk},\,\Theta_{\rm sk},\,n_{\rm sk},\,x,\,d_0,\,{\rm and}\,\theta_{\rm sk}$ represent the magnetic field, local dimensionless temperature, number density of electrons, horizontal distance from disc centre, intercept of disc surface on jet axis, and angle between jet axis and SKD surface. So for a given source
point on the SKD D$^\prime$ in Fig. \ref{fig:compo}, we compute
various moments of the intensity. B$^\prime$D$^\prime$ is the perpendicular at D$^\prime$. The differential area at D$^\prime$
is projected at P. The inner edge of SKD is $\xs$ but the lower
limit of integration for SKD moments is not $\xs$, because of the  effect of the PSD \citep{c05}. For a certain point $r$ on the jet axis, the inner edge is at
\begin{equation}
x_i (r)=\frac{r-d_0}{(r-\hs)/\xs+{\rm cot}\theta_{\rm sk}}
\label{eq:rin}    
\end{equation}
Here $\theta_{\rm sk}$ is the polar angle of the SKD surface,
ad $d_0$ is the intercept of the top surface of the SKD on the axis.

The radiative energy density, radiative flux, and radiation pressure are the frequency integrated zeroth, first, and second order moments of specific energy. The radiative moments are calculated at each point on the jet axis. The moments from the SKD on or near the jet axis are given as \citep{vkmc15}:

\begin{equation}
\mathcal{E}_{\rm sk}=\int_{x_i}^{\xo}\int_0^{2\pi} F_{\rm sk}\frac{ r d\bar{r} d\phi}{\left[\left(r-\bar{r}\rm{cot}\theta_{\rm sk}\right)^2+\bar{r}^2\right]^{3/2}},
\label{eq:skd_ez}
\end{equation}
\begin{equation}
\mathcal{F}_{\rm sk}=\int_{x_i}^{\xo}\int_0^{2\pi} F_{\rm sk}\frac{\left(r-\bar{r}\rm{cot}\theta_{\rm sk}\right) r d\bar{r} d\phi}{\left[\left(r-\bar{r}\rm{cot}\theta_{\rm sk}\right)^2+\bar{r}^2\right]^{2}},
\label{eq:skd_fz}
\end{equation}
\begin{equation}
\mathcal{P}_{\rm sk}=\int_{x_i}^{\xo}\int_0^{2\pi} F_{\rm sk}\frac{\left(r-\bar{r}\rm{cot}\theta_{sk}\right)^2 r d\bar{r} d\phi}{\left[\left(r-\bar{r}\rm{cot}\theta_{\rm sk}\right)^2+\bar{r}^2\right]^{5/2}},
\label{eq:skd_pz}
\end{equation}
where 
\begin{equation}
F_{\rm sk}=\mathcal{S}_{\rm sk}\left(\frac{u_{\rm o}\xo H_{\rm o}}{u_{\rm sk}\bar{r}H} \right)^{3(\Gamma-1)} \frac{\left(\bar{r}{\rm cos}\theta_{\rm sk}+d_0{\rm sin}\theta_{\rm sk}\right)}{\bar{r}u^2_{\rm sk}\left(\bar{r}\rm{cot}\theta_{\rm sk}+d_0\right)^2} 
\label{eq:func_sk}
\end{equation}

where $\mathcal{S}_{\rm sk}$ is a constant given as
\begin{equation}
\mathcal{S}_{\rm sk}=\frac{9.22\times10^{33}e^4\Theta_0^3\beta\sigmat\mdotsk^2}{\pi \mel^2\mp ^2c^2G^2\msol^2}
\label{eq:skd_cons}    
\end{equation}
Here $\mdotsk$ represents the accretion rate of SKD in units of Eddington accretion rate $\left(\dot{M}_{\rm Edd}=1.44\times10^{17}(M_{\rm{B}}/\msol) \, \rm{gs^{-1}}\right)$. 
It may be noted that, $u_{\rm sk}$ is the radial velocity of the accretion disc and may be estimated to be \citep[see][]{vkmc15},
\begin{eqnarray}
 u_{\rm sk}=\left[1-\frac{(x-1)x^2}{\{x^3-[(x-1)\lambda^2]\}u_t^2|_{x_0}}\right]^{1/2}; \\ \nonumber
 \mbox{ where; } (u_t)^2|_{x_0}=\left(1-\frac{1}{x_0}\right)\frac{1}{1-u^2_{0}}.\frac{x^3_0}
{x^3_0-(x_0-1)\lambda^2_0}.
\end{eqnarray}
In the above equation, $x$ is radial coordinate of the accretion disc,
$x_0$ is the outer edge and $\lambda_0$ is the specific angular momentum at the outer edge. We have considered $\lambda_0=1.7$ as a representative case. 

\begin{figure}
\includegraphics[width=\columnwidth]{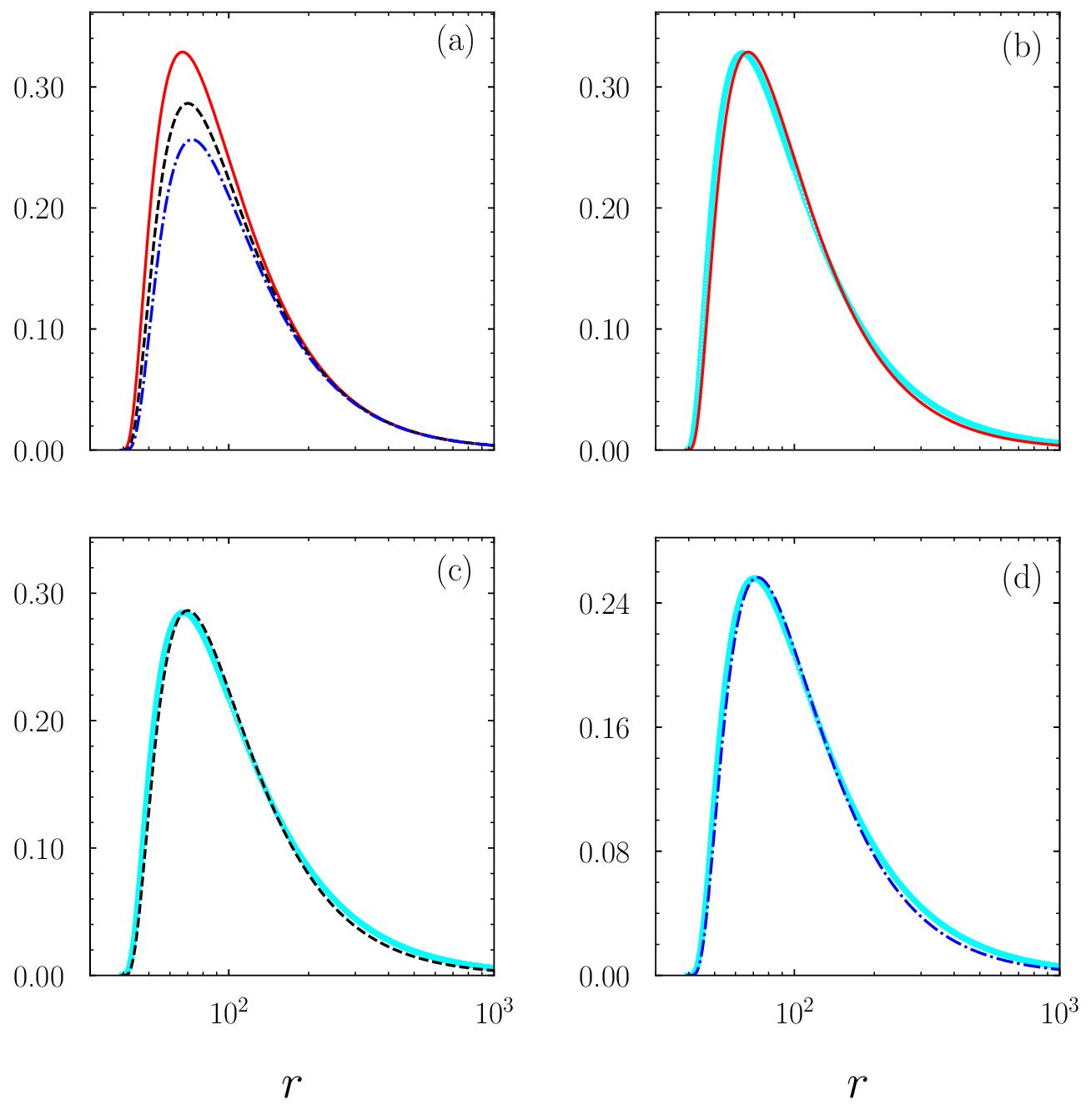}
\caption{Radiative moments for SKD. (a) Distribution of radiative moments $\mathcal{E}_{\rm sk}/S_{\rm sk}$ (solid red), $\mathcal{F}_{\rm sk}/S_{\rm sk}$ (dashed black), and $\mathcal{P}_{\rm sk}/S_{\rm sk}$ (dash-dotted blue) obtained using the numerical integration. Panels (b), (c), and (d) show a comparison between the functional forms used to mimic moments (represented by open cyan circles) and results from numerical integration for $\mathcal{E}_{\rm sk}/S_{\rm sk}$, $\mathcal{F}_{\rm sk}/S_{\rm sk}$, and $\mathcal{P}_{\rm sk}/S_{\rm sk}$ respectively.}
\label{fig:skd_mom}
\end{figure}

One has to numerically integrate the equations (\ref{eq:skd_ez}-\ref{eq:skd_pz}) to obtain the radiative moments arising out of the SKD. However in the simulation code, if one has to perform numerical integration at each time step and at each cell, then the code slows down significantly. To tackle this problem, in Appendix \ref{sec:app2} we fitted algebraic functions with the numerically obtained $\mathcal{E}_{\rm sk}$ (equation \ref{eq:eng_alb}), $\mathcal{F}_{\rm sk}$ (equation
\ref{eq:forc_alb}), and $\mathcal{P}_{\rm sk}$ (equation \ref{eq:pres_alb}). In Fig. \ref{fig:skd_mom}a, various moments
of radiation field $\mathcal{E}_{\rm sk}/S_{\rm sk}$ (solid, red),
$\mathcal{F}_{\rm sk}/S_{\rm sk}$ (dashed, black) and
$\mathcal{P}_{\rm sk}/S_{\rm sk}$ (dash-dotted, blue) by integrating
equations (\ref{eq:skd_ez}-\ref{eq:skd_pz}).
In Fig. \ref{fig:skd_mom}b we compared the numerically
integrated distribution of $\mathcal{E}_{\rm sk}/S_{\rm sk}$
(solid, red) with analytical function as in equation \ref{eq:eng_alb}
(open cyan circles). In Fig. \ref{fig:skd_mom}c, we compared
numerically integrated $\mathcal{F}_{\rm sk}/S_{\rm sk}$
(dashed, black) with the analytical function as in equation
\ref{eq:forc_alb} (open cyan circles). In Fig. \ref{fig:skd_mom}d,
we compare numerically integrated distribution of
$\mathcal{P}_{\rm sk}/S_{\rm sk}$ (dash-dotted, blue) with the
analytical expression equation \ref{eq:pres_alb} (open cyan circle).
The fitting is quite accurate. The moments are obtained for an
SKD component of the disc with $\xs=10\rg$ and $\hs=4 \xs$.

%

The PSD has much more complicated intensity profile as it involves emission from synchrotron and bremmsstrahlung processes and the inverse-Comptonization of these photons, so a proper radiative transfer treatment of the accretion disc is required to obtain the exact intensity profile from PSD which is beyond the scope of this paper hence we have made simplifying assumptions similar to \cite{cc02b} and assume the uniform intensity for PSD and is given
by
\begin{equation}
I_{\rm{ps}}=l_{\rm ps}L_{\rm Edd }/\pi A_{\rm ps}
\label{eq:psd_inten}    
\end{equation}
Where $A_{\rm ps}$ and $L_{\rm Edd }$ are surface area of PSD and Eddingnton luminosity, respectively and luminosity of PSD $l_{\rm ps}$ is in units of $L_{\rm Edd }$.\\
The radiative moments from PSD are given as 

\begin{equation}
\mathcal{E}_{\rm{ps}}=\mathcal{S}\int_{\xin}^{\xs}\int_0^{2\pi} \frac{ r\bar{r}d\bar{r} d\phi}{\left[\left(r-\bar{r}{\rm cot}\theta_{\rm ps}\right)^2+\bar{r}^2\right]^{3/2}}
\label{eq:psd_ez}
\end{equation}

\begin{equation}
\mathcal{F}_{\rm{ps}}=\mathcal{S}\int_{\xin}^{\xo}\int_0^{2\pi} \frac{ r\left(r-\bar{r}{\rm cot}\theta_{\rm ps}\right)\bar{r}d\bar{r} d\phi}{\left[\left(r-\bar{r}{\rm cot}\theta_{\rm ps}\right)^2+\bar{r}^2\right]^{2}}
\label{eq:psd_fz}
\end{equation}

\begin{equation}
\mathcal{P}_{\rm{ps}}=\mathcal{S}\int_{\xin}^{\xs}\int_0^{2\pi} \frac{r\left(r-\bar{r}{\rm cot}\theta_{\rm ps}\right)^2\bar{r}d\bar{r} d\phi}{\left[\left(r-\bar{r}{\rm cot}\theta_{\rm ps}\right)^2+\bar{r}^2\right]^{5/2}}
\label{eq:psd_pz}
\end{equation}
Where $\mathcal{S}$ is a constant given as
\begin{equation}
\mathcal{S}=\frac{1.3\times10^{38}\lpsd\sigmat}{2\pi\mel A_{\rm{ps}}G\msol}
\label{eq:psd_const}    
\end{equation}

And $\bar{r}$ is related with $r^{\prime}$ as

\begin{equation}
\bar{r}=r^{\prime}{\rm sin}\theta_{\rm ps}
\end{equation}

The moments from post shock disc admit the analytical expression,
and they are given as \citep{cc02b,cdc04},
\begin{equation}
\mathcal{E}_{\rm{ps}}=2\pi\left[\frac{(r-\xin\cot\theta_{\rm ps})}{(r-\xin\cot\theta_{\rm ps})^2+\xin^2}-\frac{(r-\xs\cot\theta_{\rm ps})}{(r-\xs\cot\theta_{\rm ps})^2+\xs^2}\right]
\label{eq:epsd_final}
\end{equation}

\begin{equation}
\mathcal{F}_{\rm{ps}}=2\pi r{\rm sin}^2\theta_{\rm ps}\left[\frac{(r-2\xin\cot\theta_{\rm ps})}{(r-\xin\cot\theta_{\rm ps})^2+\xs^2}-\frac{(r-2\xs\cot\theta_{\rm ps})}{(r-\xs\cot\theta_{\rm ps})^2+\xs^2}\right]
\label{eq:fpsd_final}
\end{equation}

\begin{equation}
\mathcal{P}_{\rm{ps}}=2\pi\left[\frac{(r-\xin\cot\theta_{\rm ps})^3}{3\left[(r-\xin\cot\theta_{\rm ps})^2+\xin^2\right]^{3/2}}-\frac{(r-\xs\cot\theta_{\rm ps})^3}{3\left[(r-\xs\cot\theta_{\rm ps})^2+\xs^2\right]^{3/2}}\right]
\label{eq:prepsd_final}
\end{equation}

In principle, the accretion rate of SKD ($\dot m_{\rm sk}$) controls the spectral state of accretion disc, hence the luminosity of PSD ($l_{\rm ps}$) and position of shock can be calculated using the spectral modeling but in this paper, we have supplied the $l_{\rm ps}$, $\xs$ and $\dot m_{\rm sk}$ as free parameters for simplicity. 
It may be noted, the expressions of constants $S$ and $S_{\rm sk}$ (equations \ref{eq:psd_const} and \ref{eq:skd_cons}) are in the geometric unit system where $2G=M_{\rm \small B}=c=1$ ($M_{\rm \small B}$ is the mass of black hole and $G$ is the gravitational constant). Hence in this unit system unit of length is the Schwarzschild radius ($\rg= 2GM_{\rm \small B}/c^2$) and unit of time is $t_{\rm g}=2GM_{\rm \small B}/c^3$.
In physical units these would be $\rg=3\times10^5~m_{\rm \small B}$ cm
and $t_{\rm g}=10^{-5}~m_{\rm \small B}$ s, where the BH mass in units
of solar mass is given by $m_{\rm \small B}=M_{\rm \small B}/M_\odot$.

The radiative moment at any field point P (refer Fig. \ref{fig:compo}) is the sum of the contribution from both PSD and SKD, so
\begin{eqnarray}
 \mathcal{E}=\mathcal{E}_{\rm ps}+\mathcal{E}_{\rm sk}, \\
 \mathcal{F}=\mathcal{F}_{\rm ps}+\mathcal{F}_{\rm sk}, \\
 \& ~~~ \mathcal{P}=\mathcal{P}_{\rm ps}+\mathcal{P}_{\rm sk}.
\end{eqnarray}

\begin{figure}
\includegraphics[width=\columnwidth]{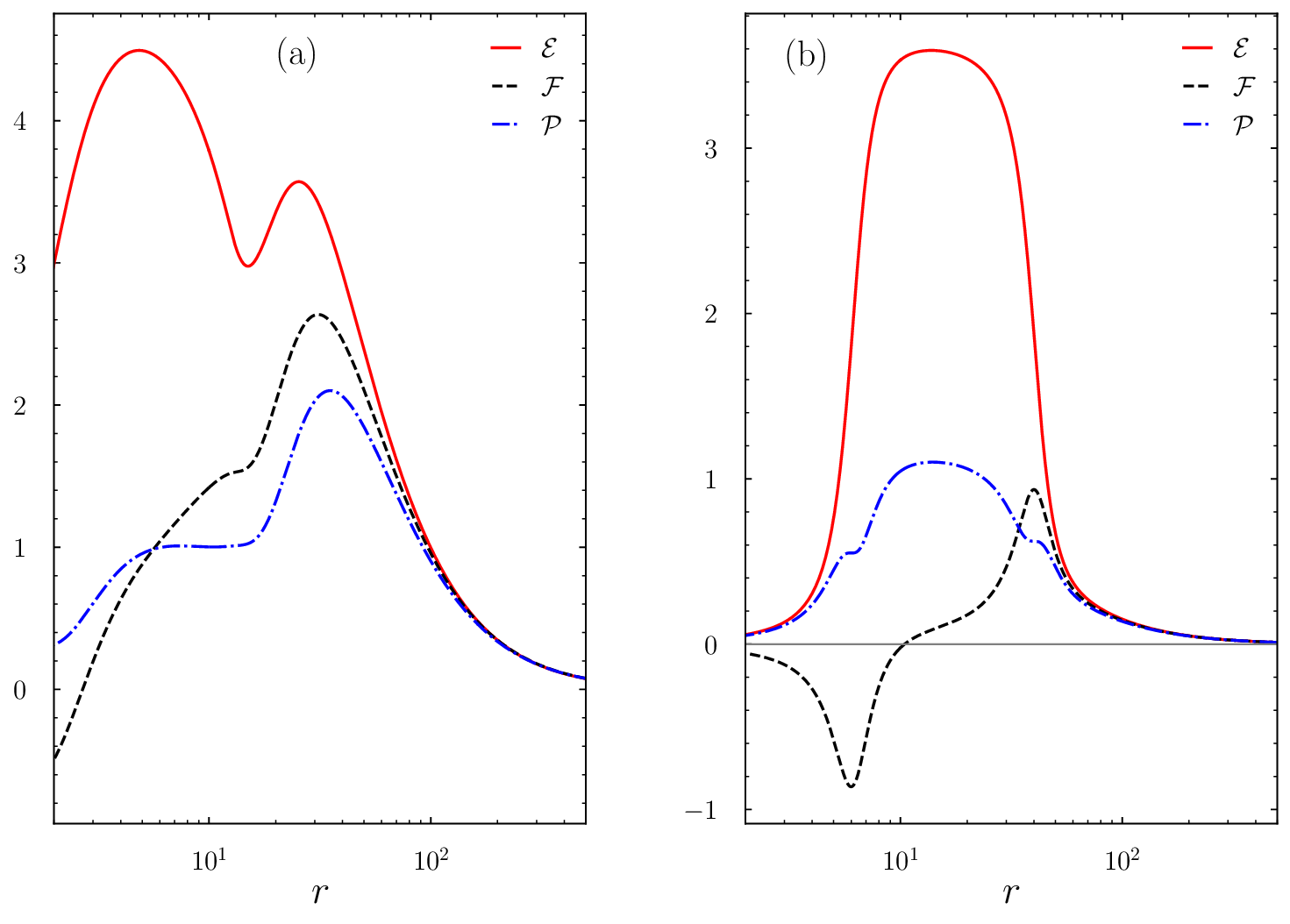}
\caption{Distribution of radiative moments- $\mathcal{E}$ (solid red line), $\mathcal{F}$ (dashed black line), and $\mathcal{F}$ (dashed dotted blue line) for two different disc configurations, $\hs=\xs$ (panel (a)) and $\hs=4\xs$ (panel (b))}
\label{fig:combined_moments}
\end{figure}

In Figs. \ref{fig:combined_moments}a,b, we have plotted the radiative moments $\mathcal{E},\, \mathcal{F},\, \mathcal{P}$ for $l_{\rm ps}=0.2$ and $\dot m=5$. In the left panel
Fig. \ref{fig:combined_moments}a, location of shock is at taken as $\xs=13.2r_g$ and shock height is $\hs=\xs$. We can clearly distinguish two peaks in the distribution of the moments because the radiative moments from different part of the disc peak at different locations. In regions closer to the central object, the moments from PSD dominate in the overall contribution, while at the larger distances $r>20$ the radiation from SKD starts to dominate. This can result in multistage acceleration of the jet.  
On the other hand Fig. \ref{fig:combined_moments}b, shows the distribution of moments for a post shock region which is geometrically thicker with shock height $\hs=4\xs$.
When the PSD is thick, within the funnel region the jet material will see some fraction of radiative flux coming towards it hence the radiation flux inside the funnel is negative as shown in Fig. \ref{fig:combined_moments}-b. The incoming flux opposes the outflowing jet material which opens up the possibility of multiple sonic points and shocks in the jet-flow as the supersonic material is being slowed down by the radiation \citep{vc18}.  


\section{Numerical setup}
\label{sec:method}

\subsection{Simulation Setup}
We simulate the jets in the length scale upto $1000r_g$. The speed of light is taken as the unit of velocity in the code. 
We have employed an outflow boundary condition at outer boundary and continuous boundary condition at the injection cell using the ghost cells. We first obtain the steady state jets by solving
equations of motion (equations \ref{eq:dvdr},\ref{eq:dthetadr}) and solve them as is described in section \ref{sec:simulationcode}. The associated accretion disc is in steady state, therefore, the radiation field is also in steady state. We inject the numerical simulation code (section \ref{subsec:code}) at the jet base
with flow variables from the semi-analytical, steady state jets. We then compare the simulation with the steady state solutions, and find out how well the time dependent code
regenerates the steady state solution. 

Since this is a time dependent study, we would like to study how the time dependence of the accretion disc affect the jet solution. However, we are not imposing any time dependence on the jet base to make it a time varying jet. 
Instead we invoke a time varying disc, where the inner part of the disc (read PSD) is in motion. It may be noted that, the inner part of the disc produces high energy photons and behaves as the illusive corona related to accretion discs.
Many authors have identified oscillation of inner part of the accretion discs as the
origin of quasi-periodic oscillation or QPOs \citep{nandi12}. 

\subsection{The disc as seen by an inertial observer in the jet}

Regeneration of the steady state bipolar outflow is devised to check
the performance of the code and how well the numerical code can capture
the steady state theoretical solution. However, the steady state scenario is also used in this paper as preprocessed initial jet structure.
The time dependence is imposed on the steady jet through the resulting time dependent radiation field. Here the accretion disc which produces
the radiation field is not part of the computational domain.
The inner part of the accretion disc is in quasi periodic oscillation. We approximate this with a sine
function of the radius of the outer edge of PSD or $\xs$.
So when $\xs$ decreases, PSD contracts but SKD expands.
It may be noted that the expressions of $\mathcal{E}$,
$\mathcal{F}$ and $\mathcal{P}$ depend on $\xs$, $x_i$
(equations \ref{eq:epsd_final}-\ref{eq:prepsd_final} and
\ref{eq:eng_alb}, \ref{eq:forc_alb}, \ref{eq:pres_alb}).
Therefore at a location $r$ on the jet, if $\xs$ is time dependent, then the radiation field from the SKD and PSD, will also vary in time.

\begin{figure}
\includegraphics[width=11.0cm]{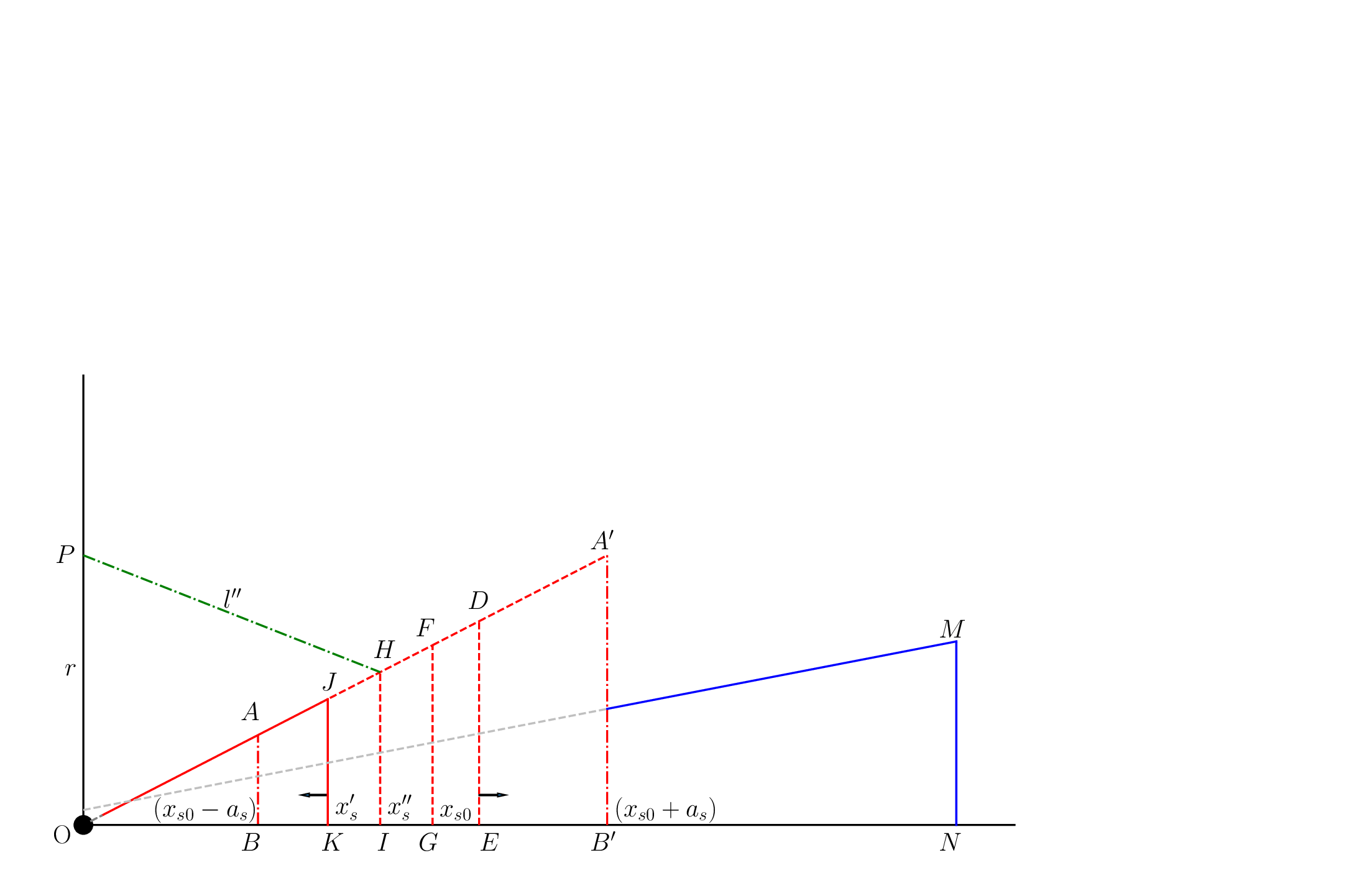}
\caption{The retarded position of shock ($x^{\prime\prime}$) as measured by an observer $P$ on jet axis. $x_0$ is the initial position of the shock and $x^{\prime}$ represent the actual position of the shock at time $t$.}
\label{fig:schematic_shock}
\end{figure}

The situation is illustrated in Fig. \ref{fig:schematic_shock}. Let $x_{\rm s0}$ be the mean position of the outer edge of the PSD (the surface represented as FG) and let $a_{\rm s}$
be the amplitude of oscillation (at AB \& A$^\prime$B$^\prime$). The frequency of oscillation be $f_{\rm s}$. At time $t$, let the outer edge of PSD is at $\xs^\prime$ (say, i. e., at JK).
It may be noted that
the information about any change in the accretion disc configuration does not reach the jet axis instantaneously as the photons emitted from the disc travel with the speed of light and take a finite amount of time to reach the jet axis. Hence at any epoch, the disc configuration for different points on jet will not be same. So the light ray from the outer edge of the PSD, that reaches the jet at point P in time $t$, should
be emitted when the outer edge of PSD was at an earlier position
$\xs^{\prime \prime}$ (HI).
The general expression of the location of the outer edge of
oscillating PSD is
\begin{equation}
\xs^\prime=x_{\rm s0}+a_{\rm s}\,{\rm sin} (2\pi f_{\rm s} t)
\label{eq:shockpos}    
\end{equation}

Following Fig. \ref{fig:schematic_shock}, let us compute
the 
radiation field at P and at time $t$, in the epoch when the shock is moving inward. The radiation from the outer
edge of PSD that reaches point P at time $t$, is at $\xs^{\prime \prime}$. 
So the time taken for light to reach from H to P be $\delta t$ is the same time in which shock has moved from $\xs^{\prime\prime}$ to $\xs^{\prime}$, as shown in Fig \ref{fig:schematic_shock}.   
\begin{equation}
\delta t=l^{\prime\prime}/c=\sqrt{{\xs^{\prime\prime}}^2+(r-\hs^{\prime\prime})^2}  \mbox{;  in units of } c=1
\label{eq:delta_t}
\end{equation}
The instantaneous velocity of outer edge of PSD is $v_{\rm s}$ and can be obtained from equation \ref{eq:shockpos}
\begin{equation}
 v_{\rm s}=a_{\rm s}2\pi f_{\rm s} {\rm cos}(2\pi f_s t)
 \label{eq:instvelsok}
\end{equation}
The position of shock $\xs$ is updated after each time interval $dt$. The interval $dt$ is determined by the TVD code itself.
Assuming a small $dt$ we can write
\begin{equation}
\xs^{\prime \prime}-\xs^{\prime}=s dt  
\label{eq:pos_diff}
\end{equation}
Where $s=0.5(v_{{\rm s},\,t}+v_{{\rm s},\,t+dt})$ is the average velocity in the time interval between $t$ to $t+dt$.
Using equations (\ref{eq:delta_t}) and (\ref{eq:pos_diff}), we can write
\begin{equation}
(\xs^{\prime\prime}-\xs^{\prime})^2=s^2\left[{\xs^{\prime\prime}}^2+(r-h_s^{\prime\prime})^2\right]    
\label{eq:ret_pos}
\end{equation}
For each $r$, equation (\ref{eq:ret_pos}) is solved at every time step to obtain the shock location as seen by the observer at $r$, and the radiative moments for corresponding shock location are calculated. 

\section{Results}
\label{sec:results}
%
%
%
\subsection{Code verification}
The purely hydrodynamic codes in Cartesian coordinates are tested
against initial value problem like the exact solution of shock-tube
problem. In the following, we test the code in the spherical coordinate system with the analytical steady state
bipolar radiatively driven outflow solutions. It may be noted, purely thermally driven radial outflows are regenerated well. It may also be commented that the outflow requires higher resolution, compared to the
accretion solution.

We use the steady state solutions as test problem for our simulation code as the exact solutions can be obtained for steady state case.  We obtain the solutions for electron-proton jets i.e. $\xi=1.0$. 

\begin{figure}
\includegraphics[width=\columnwidth]{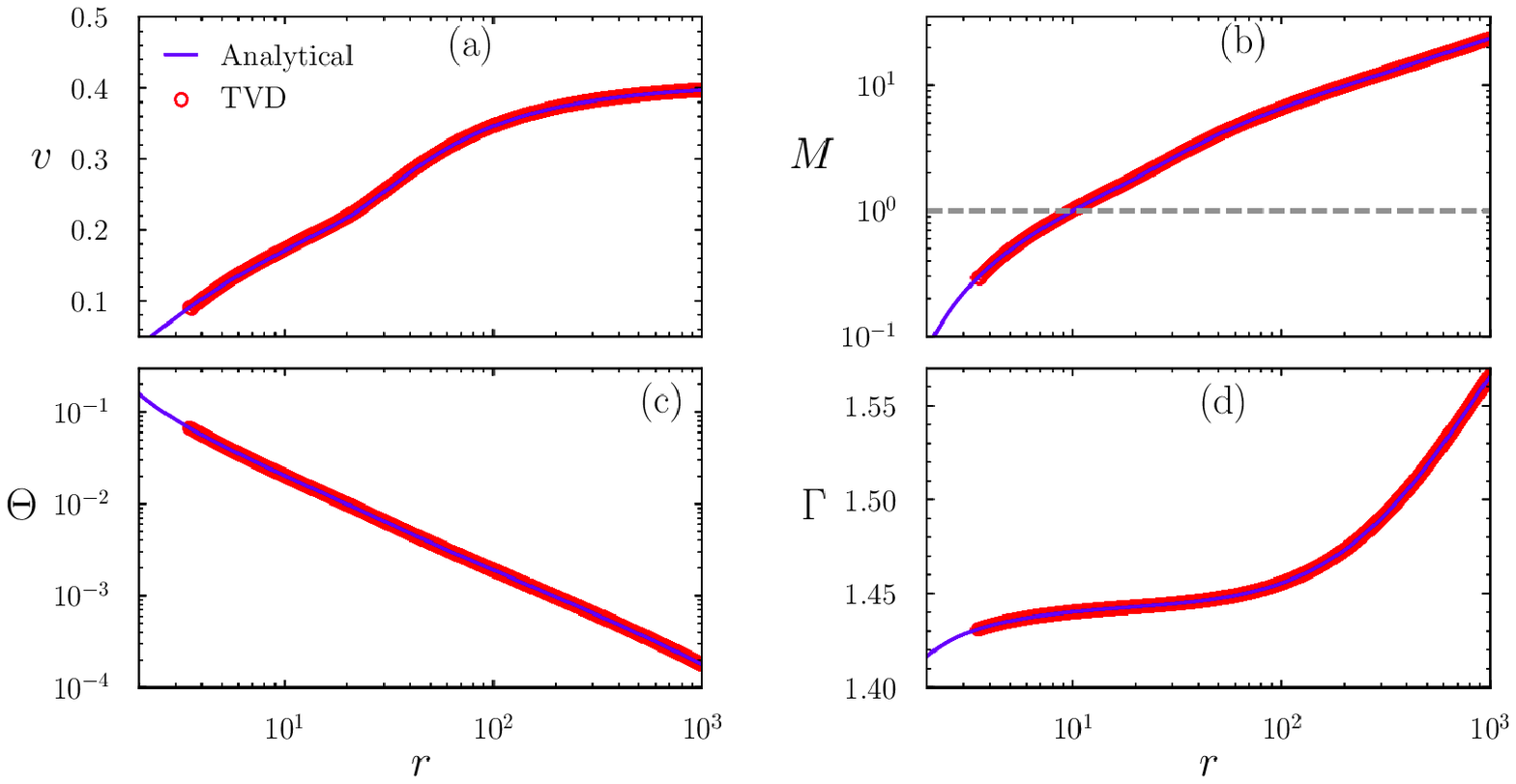}
\caption{Comparison between analytical solution (solid blue line) and solution obtained from simulation code (red open circles) for the disc configuration $\hs=\xs$. Variation of (a) jet velocity, (b) Mach number, (c) dimensionless temperature (d) adiabatic index are plotted along the direction of propagation of jet.}
\label{fig:comp_thin_corona}
\end{figure}

In Fig. (\ref{fig:comp_thin_corona}a-d) we have shown the comparison of the exact solutions and solutions obtained by simulation code for a steady state jet. Various panels show the evolution of different flow variables like jet velocity $v$ (Fig. \ref{fig:comp_thin_corona}a), Mach number $M$ (Fig. \ref{fig:comp_thin_corona} b), $\Theta$ (Fig. \ref{fig:comp_thin_corona} c) and adiabatic index (Fig. \ref{fig:comp_thin_corona}d). In this case have assumed a PSD with shock height $\hs=\xs$. The disc parameters to obtain the radiation field are given as
\begin{equation}
\xs=13.2r_g,\, l_{\rm{ps}}=0.2, \, \dot{m}_{\rm{sk}}=5.0 
\label{eq:disc_par_thin}    
\end{equation}

The sonic point of the flow is at $r_c=10.0 r_g$. To verify the simulation code we take the injection parameters from the analytical solution. The injection parameters are taken as $\vin=0.0906,\, \thin= 0.0669$ at $\rin=3.5$. We have divided the computational domain in 6000 uniform cells.  
Comparison of various flow variables like $v$, $M$, $\Theta$, and adiabatic index $\Gamma$ with the analytical solutions shows that the TVD code generates solution with a very good accuracy. 
Figure {\ref{fig:comp_thin_corona}}-(c) shows that the jet temperature reduces by four order of magnitudes consequently the adiabatic index also shows a transition from a thermally relativistic value (1.43) to a non-relativistic value (1.54), which highlights the importance of using an EoS with variable adiabatic index. \\

\begin{figure}
\includegraphics[width=\columnwidth]{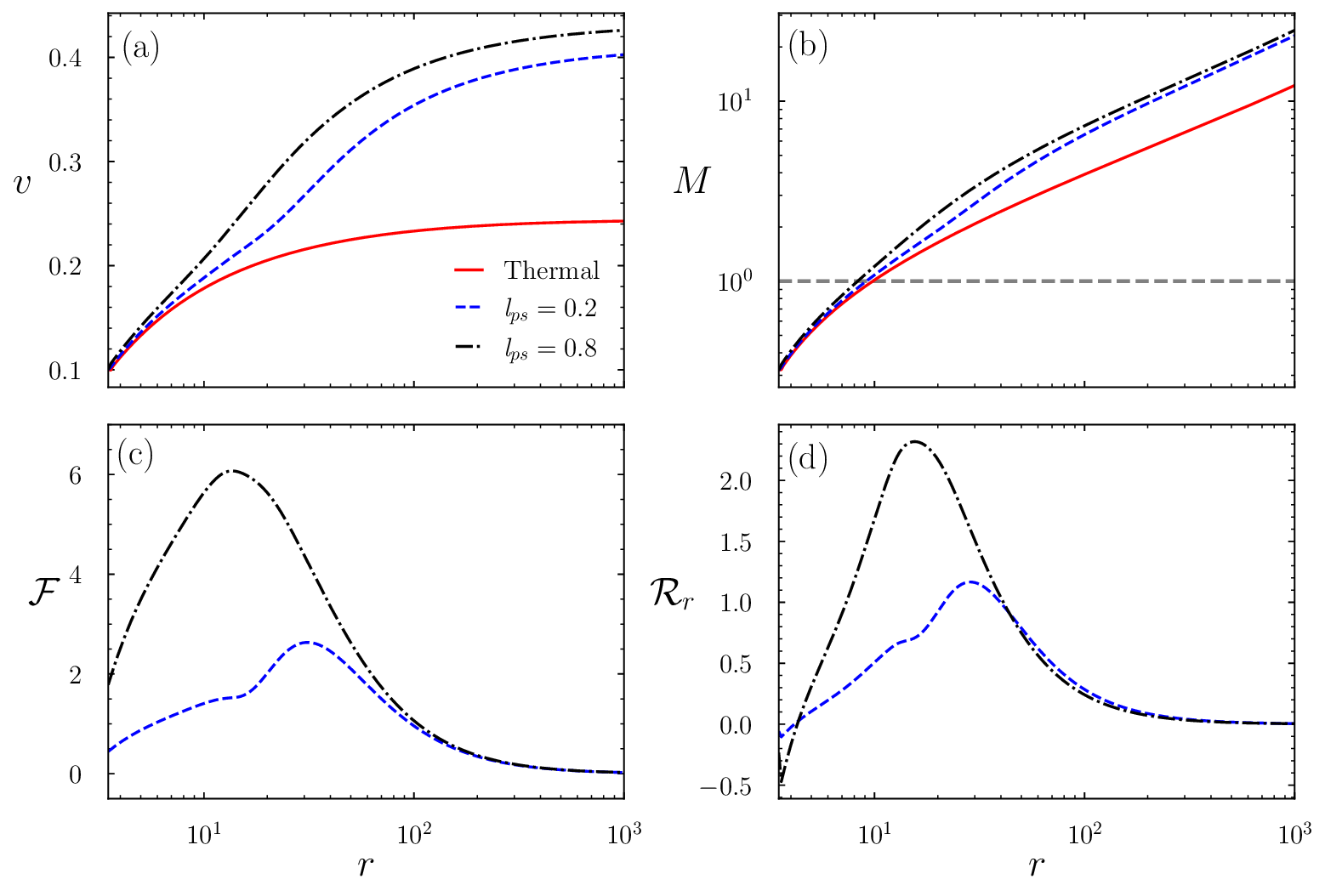}
\caption{Jet solutions driven with post-shock luminosity $l_{ps}=0.2$ (dashed blue), $l_{ps}=0.8$ (dash-dotted black), and purely thermal driven (solid red) for the disc configuration $\hs=\xs$. Injection parameters are same for all solutions. Different panels show the variation of (a) jet velocity , (b) Mach number, (c) $\mathcal{F}$, and (d) radiative acceleration.}
\label{fig:comp_diif_lum_thin}
\end{figure}

In Fig. (\ref{fig:comp_diif_lum_thin}) we compare the jet solutions driven by radiation arising out of PSD with luminosity $l_{\rm{ps}}=0.2$ (dashed blue) and $l_{\rm{ps}}=0.8$ (dash-dotted black) with a purely thermal driven jet (solid red line). The injection parameters are kept same for all three   
solution  $\vin=0.1002,\, \thin= 0.0684$ at $\rin=3.5$. With these parameters the thermally driven jet becomes supersonic at $r_c\simeq 9.8$ and reaches up to the speed $v=0.24c$. The effect of radiation in accelerating the jet is evident from fig \ref{fig:comp_diif_lum_thin}(a). The increment in the luminosity of the disc results in higher radiation flux and higher value of radiative contribution which is clearly visible from panels (c) and (d). 
The jet achieves the speed upto $v=0.40c$ and $v=0.43c$ for $l_{\rm{ps}}=0.2$ and $l_{\rm{ps}}=0.8$. It implies that radiative acceleration can increase the terminal speed by $66.6\%$ and $79.2\%$ respectively, over pure thermally driven wind. In Fig. (\ref{fig:comp_diif_lum_thin}b) we have plotted the variation of Mach number ($M$), dashed grey line represents $M=1$ line. The acceleration brings sonic point towards the jet base.

\begin{figure}
\includegraphics[width=\columnwidth]{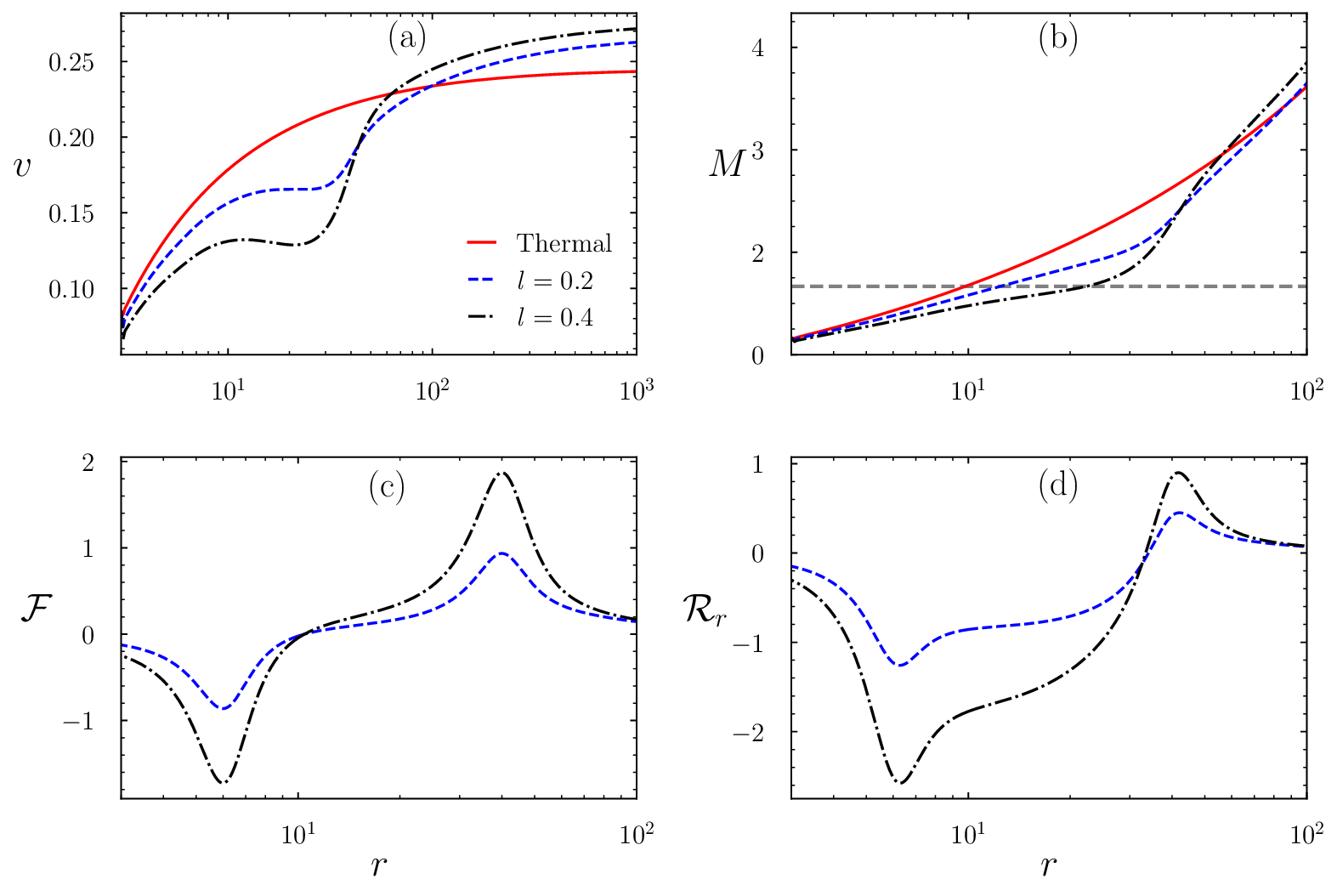}
\caption{Jet solutions driven with post-shock luminosity $l_{\rm ps}=0.2$ (dashed blue), $l_{\rm ps}=0.4$ (dash-dotted black), and purely thermal driven (solid red). Injection parameters are same for all solutions. Different panels show the variation of (a) jet velocity , (b) Mach number, (c) $\mathcal{F}$, and (d) $\mathcal{R}_r$.}
\label{fig:comp_diif_lum}
\end{figure}

In Fig. (\ref{fig:comp_diif_lum}), we study the effect of geometrically thick PSD i. e., when $\hs=4\xs$. It may be noted that, for oscillating discs the inner PSD region may become geometrically thick as $\xs$
approaches the central object
\citep{lcksr16}. We compare the jet solutions corresponding to different disc luminosity for a geometrically thick PSD. The injection parameters are taken as $\vin=0.085$ and $\thin=0.0838$ at $\rin=3.0r_g$.
The disc parameters that produce the radiation field is 
\begin{equation}
\xs=10r_g,\, \dot{m}_{\rm{sk}}=5.0 
\label{eq:disc_par_thick}    
\end{equation}
The solid red line represents the jet solution which is thermally driven with these injection parameters and the sonic point obtained is at $r_c=10r_g$. The radiatively driven jets are represented by dashed-blue
($l_{\rm ps}=0.2$) and dash-dotted-black ($l_{\rm ps}=0.4$) curves, respectively. It is interesting to note that the sonic point behaviour in Fig. \ref{fig:comp_diif_lum} is opposite to the previous
one. For moderately thick PSD
(e. g., Fig. \ref{fig:comp_diif_lum_thin}), the sonic point decreases with the increase of $l_{\rm ps}$. This is expected,
since the jet is getting accelerated, it is crossing the sonic barrier
at a shorter distance from the base. However, Fig. (\ref{fig:comp_diif_lum}a, b) show that jet is ending up with higher
terminal speed with the increase of $l_{\rm ps}$, but $r_c$ increases. If the PSD is geometrically thick then, the radiation `looks down' on the jet axis upto a much higher distance so the radiation actually decelerates the jet.
as a result the jet travels a longer distance to cross the sonic barrier.  
Panel (c) shows that the increase in the luminosity makes $\mathcal{F}<0$. Therefore, not only $\mathcal{E},\,\mathcal{P}$ combines
with the $v$ to decelerate the jet, but $\mathcal{F}$ too opposes the forward expansion of the jet. In panel (d) we plot the radiative contribution term $\mathcal{R}_r$ within the funnel region and show that indeed there is significant
deceleration near the jet base. $\mathcal{F}<0$ for $r<10$, but $\mathcal{R}_r < 0$ for $r\lsim 30$. 
Consequently, the jet driven by a disc with higher luminosity is slower in the region $r\lsim 30$ which can be clearly seen in panel (a). The sonic point moves away from the jet base as the luminosity of PSD increases, while in the case of a geometrically moderate PSD ($\hs=\xs$) (see Fig. \ref{fig:comp_diif_lum_thin}b) sonic point comes towards the jet base. However, $\mathcal{R}_r$ shows that at $r>30$, higher $l_{\rm ps}$ produces more acceleration and the resulting terminal speeds are higher.
So for discs with geometrically thick PSDs, higher $l_{\rm ps}$ produces jets with higher
terminal speed, but the sonic points are located further away from the jet base.

\begin{figure*}
\includegraphics[width=18cm, height=14cm]{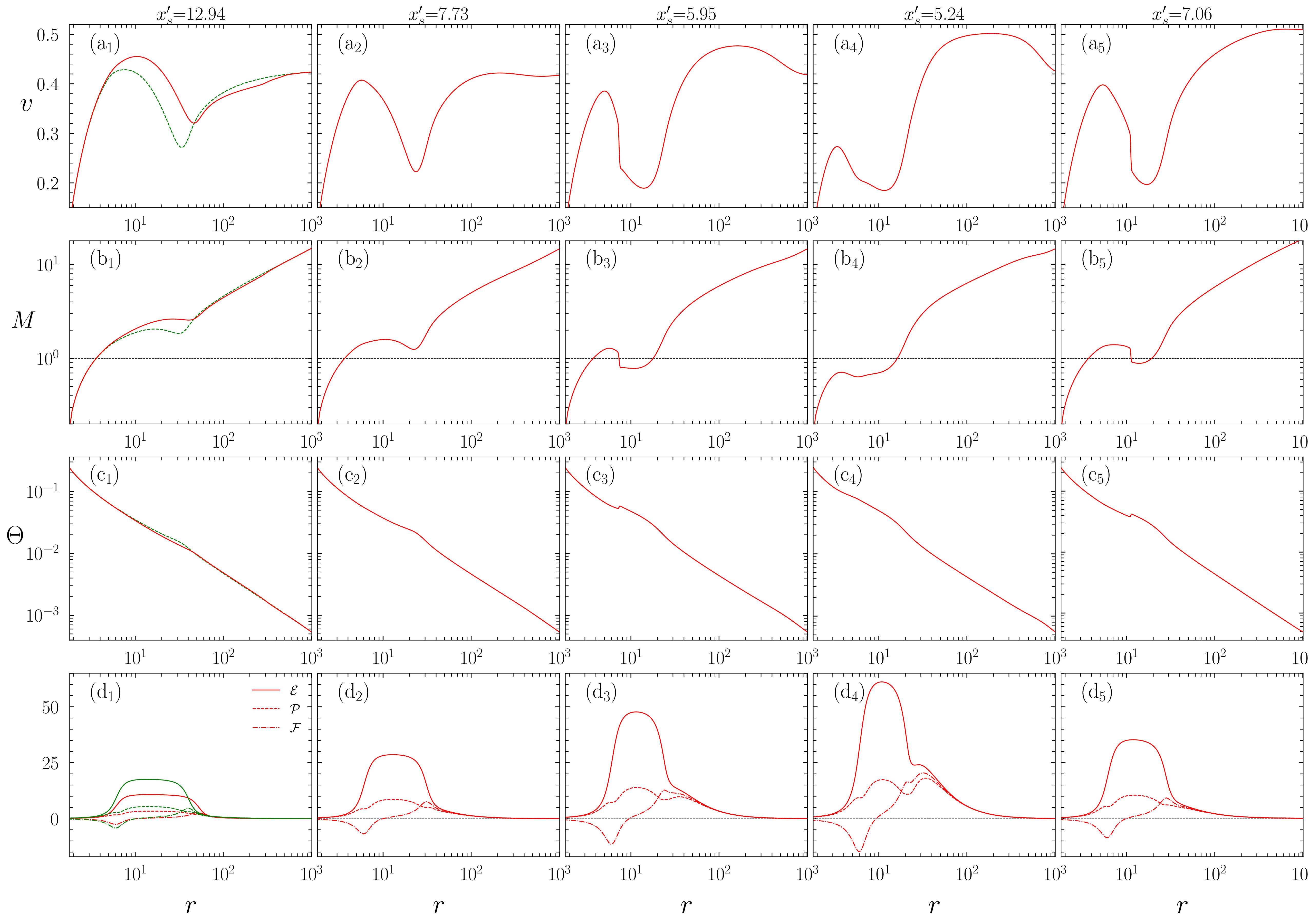}
\caption{The time dependent jet solution for different accretion disc shock locations as marked on the top of each column. Panels (${\rm a_1-a_5}$), (${\rm b_1-b_5}$) and (c$_1$---c$_5$) show the jet velocity, Mach number and $\Theta$, respectively. The radiative moments $\mathcal{E}$ (solid red), $\mathcal{P}$ (dashed red), and $\mathcal{F}$ (dash-dotted red) are plotted in panels (${\rm d_1-d_5}$). The steady state results are plotted with green colour. The radiative moments corresponding to the steady state configuration ($\xs=10$) are plotted with green color in panel (${\rm d_1}$).}
\label{fig:model_1}
\end{figure*}

\subsection{Time dependent solutions}
In this section we show the effect of inner accretion disc oscillation on the jet solutions. As described earlier,
the outer edge of the PSD i. e. $\xs$ oscillates between the positions $x_{{\rm s} 0}-a_{\rm s}$ and $x_{{\rm s}0}+a_{\rm s}$, where $a_{\rm s}$ is the amplitude of oscillation and 
$x_{{\rm s} 0}$ is the mean position of the shock (see, equation
\ref{eq:shockpos}).

\subsubsection{Model 1}
In the first time dependent model, we assume a geometrically thick PSD i.e., the shock height is given as $\hs=4\xs$. We assume that the semi-vertical angle of the inner edge of PSD remains constant throughout the oscillation of $\xs$. In Fig. \ref{fig:model_1} we have shown the variation of jet velocity ($v$), Mach number ($M$), $\Theta$ and radiative moments in panels ${\rm (a_1)-(a_5)}$, ${\rm (b_1)-(b_5)}$, ${\rm (c_1)-(c_5)}$ and (d$_1$)---
(d$_5$), respectively at different time steps and $\xs^\prime$ corresponding to those time intervals is marked on the top of each column. The injection parameters are given as 
\begin{equation}
\vin=0.108, \, \thin=0.264,\, \rin=1.6
\label{eq:inject_model_1}    
\end{equation}
First we generate the steady state solution corresponding to the shock location $\xs=10.0$ and $l_{\rm ps}=1.0$ and $\dot m_{\rm sk}=10.0$ with these injection parameters. 
The variation of velocity, Mach number and distribution of radiative moments for the steady state solution are plotted in panels ${\rm (a_1)}$, ${\rm (b_1)}$, and ${\rm (c_1)}$ with green colour.
We start the simulation with $x_{\rm s0}=\xs=10.0~\rg$. Once the steady state is achieved $\xs$ starts to oscillate with amplitude $a_{\rm s}=5~\rg$ and period $T=10^4 t_{\rm g}$. Therefore $\xs$ varies between  $x^{\prime}_{{\rm s}~{\rm min}}=5$ and $x^{\prime}_{{\rm s}~{\rm max}}=15$ in a time period of $T$. The position of shock ($\xs^{\prime\prime}$) as seen by any point $r$ on the jet axis at time $t$ is calculated using equation (\ref{eq:ret_pos}), the solution of which
turns out to be

\begin{figure*}
\includegraphics[width=18cm, height=16cm]{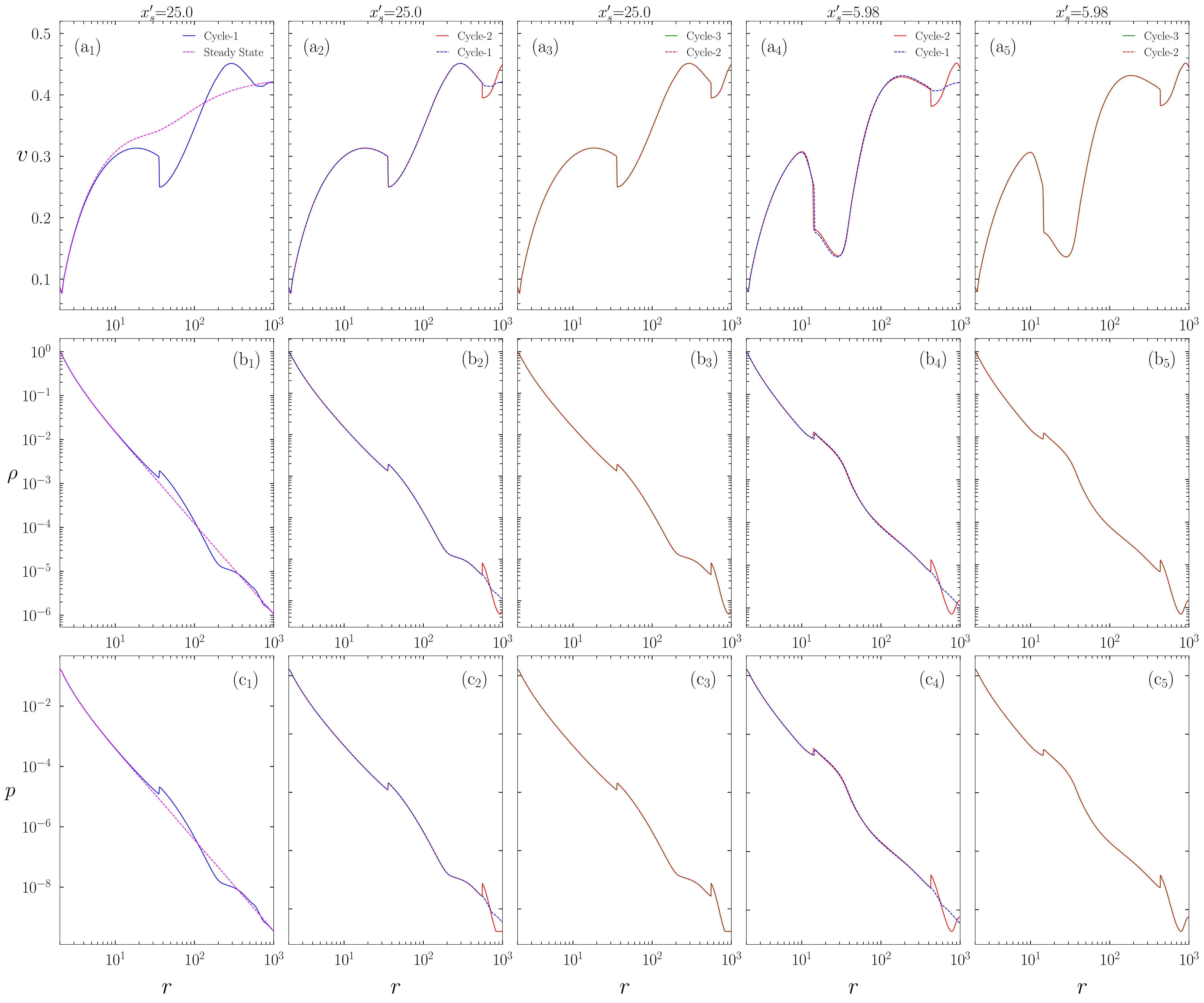}
\caption{The jet solutions in different cycles of accretion disc oscillation for disc model-2. The velocity, density (in arbitrary units), and pressure (in arbitrary units) profiles are plotted in panels $\rm a_1- a_5$, $\rm b_1- b_5$ and $\rm c_1- c_5$, respectively. The steady state solution is plotted with magenta color.}
\label{fig:model_2_cycles}
\end{figure*}

\begin{equation}
\xs^{\prime\prime}=\frac{-b\pm\sqrt{b^2-4ac}}{2a}
\label{eq:xs2_mod1}
\end{equation}
In equation (\ref{eq:xs2_mod1}), $a=1-17s^2, \, b=-(2\xs^{\prime}-8r~s^2), \, c={\xs^{\prime}}^2-s^2r^2$ and plus (minus) sign is taken when shock moves inwards (outwards). 

In Fig. (\ref{fig:model_1}) from top to bottom, we plot $v$ (a$_1$---a$_5$), $M$ (b$_1$---b$_5$), $\Theta$ (c$_1$---c$_5$) and radiative moments (d$_1$---
d$_5$). While left to right, we plot jet solutions at various times
as the shock moves inwards till the fourth column ($\xs^\prime ~ \rightarrow 12.94 \rg$---$5.24\rg$) and then
in the fifth column it starts to expand. In Figs. (\ref{fig:model_1}a$_1$,
b$_1$ c$_1$ and d$_1$), the curves in green colour show the steady state values. 
When the outer edge of PSD $\xs^\prime >x_{\rm s0}$   
the radiative moments become weaker because the area of PSD increases resulting in a lower intensity.
Hence the radiative deceleration inside the funnel reduces, so the $v$ (red)
distribution is higher than the steady state value (green). 
However, the outer part of jet ($r>50$) remains unchanged because the information about the change in the accretion disc has not reached beyond that point. The comparison of radiative moments ($\mathcal{E}$ red-solid,
$\mathcal{F}$ red-dot-dashed, \& $\mathcal{P}$ red-dashed) for a disc with $\xs^\prime =12.94$ and the steady state value (green) are plotted in
Fig. \ref{fig:model_1}.
The radiative moments become stronger when $\xs^\prime$ starts to move towards the central object, such that $\mathcal{E}$ and $\mathcal{P}$ is larger than the steady state values, interestingly $\mathcal{F}<0$ close to the jet base (see Fig. (\ref{fig:model_1}${\rm (c_2)}$). The jet still expands due to the thermal
gradient force, however, the radiative moments slows down the jet fluid inside the funnel and the jet velocity decreases. The radiative deceleration keeps on increasing as $\xs^\prime$ decreases. Such that at $\xs^\prime=5.95$ the radiative moments can drive a shock in the jet at
the location $r_{\rm js} = 7.2$. Panels ${\rm (a_3)}$, ${\rm (b_3)}$ and (c$_3$) show the velocity, Mach number and $\Theta$ profile for a shocked solution. The Mach number jumps from supersonic to subsonic value. The dotted black line represents $M=1$ line and the intersection of this line with jet solution shows the positions of sonic points of the flow. And it is quite clear the sonic point location changes as we move from the left to the right columns, to the extent that the shocked jet has two physical sonic points (Fig. \ref{fig:model_1}b$_3$). In panels ${\rm (a_4)}$, ${\rm (b_4)}$ and (c$_4$) we have plotted the $v$, $M$
and $\Theta$ of the jet for accretion disc with $\xs=5.24$. Radiative
moments are so intense (Fig. \ref{fig:model_1}d$_4$) such that the radiative deceleration is very high, so that it does not allow the flow to become supersonic within the funnel, as a result the sonic point in the jet forms at $r_c \sim 20$. After reaching its lower bound ($\xs^\prime =5$, in this case) the $\xs^\prime$ starts to move out which again gradually reduces the magnitude of moments and the jet velocity starts to increase in the funnel region of PSD and the jet-shock reappears at $r_{\rm js} = 11$.
Therefore as the PSD of the accretion disc oscillates, we create a variable jet as well as, jet shock appears and disappears, due to the intricacies
of the interaction of the jet material and the radiative moments from the accretion disc.

\begin{figure*}
 \includegraphics[width=15cm, height=11cm]{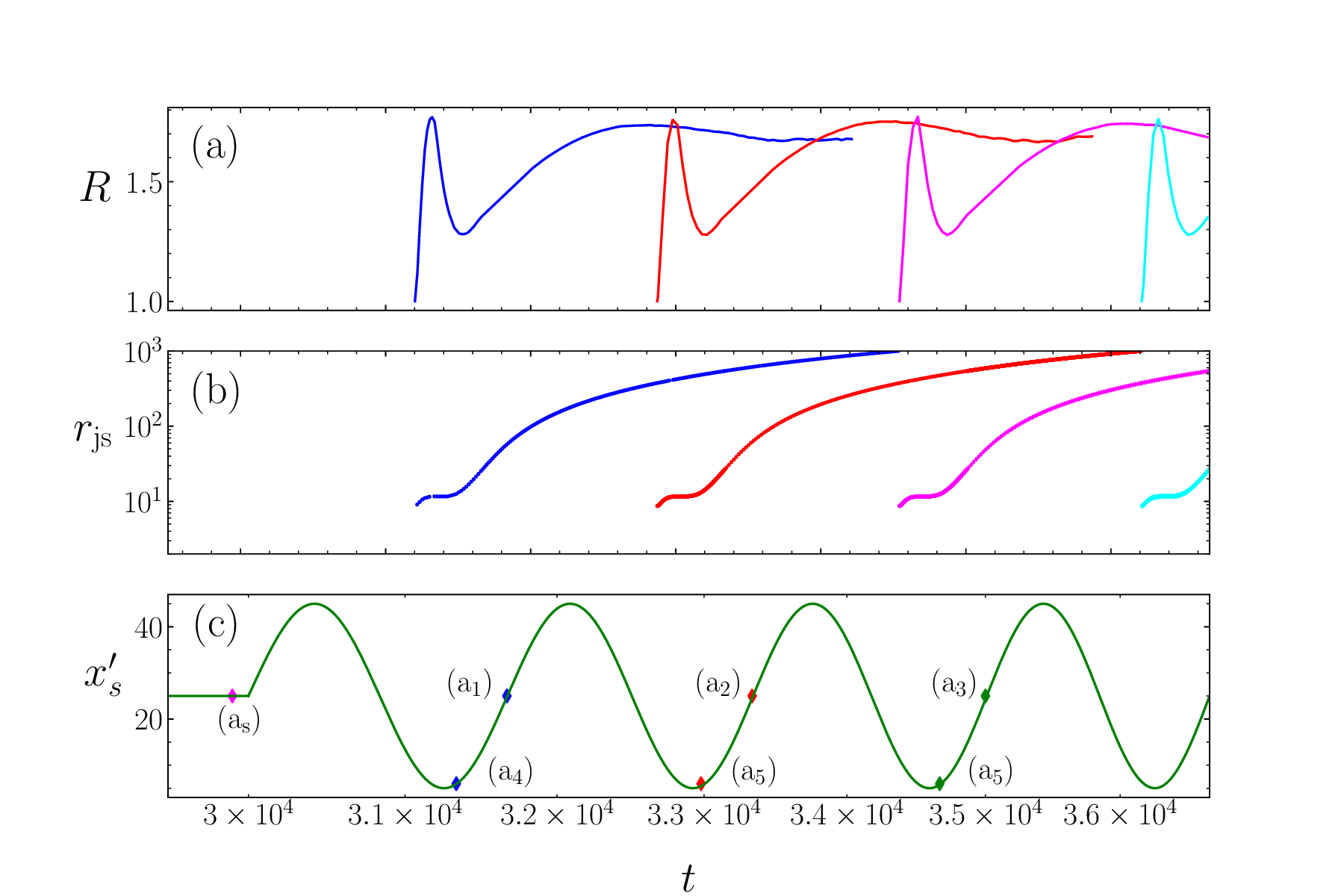}
 \caption{Variation of (a)compression ratio $R$ of the jet shock, (b) location of the jet shock
 $r_{\rm js}$ and (c) size of the PSD $\xs^\prime$ as a function of time $t$. The equilibrium PSD location in the disc is $x_{{\rm s}0}=25$.
 The injection parameters and disc luminosities are for Model 2.}
 \label{fig:jetsok}
\end{figure*}

\subsubsection{Model 2}
In this model, we assume that the height of the outer edge of PSD i. e.,
$\hs$, remains constant during the oscillation, so the PSD is geometrically thin i. e., $\hs < \xs$ as the PSD expands.
However, as the PSD contracts then $\hs > \xs$, which can generate a jet solution with multiple sonic points. The injection parameters are given as

\begin{equation}
v_{\rm in}=0.086, \, \Theta_{\rm in}=0.17,\, r_{\rm in}=2.0
\label{eq:inject_model_2}    
\end{equation}
And the PSD luminosity is $l_{\rm ps}=0.6$ and $\dot m_{sk}=10.0$. The initial value of $x_{\rm s 0}=25.0$ and the amplitude of oscillation is 20. The time period of oscillation is $T=1.67\times 10^3 t_{\rm g}$. The height of PSD is taken to be $\hs=40$. The results for this model are plotted in Fig. \ref{fig:model_2_cycles}. As before we inject with the jet-base values (equation \ref{eq:inject_model_2}) from the steady state analytical solution. We allow the simulation to settle into a steady state jet, where the accretion disc is for $\xs=25$ solution (magenta-dotted curves) for $v$ in Fig. \ref{fig:model_2_cycles}a$_1$;
 $\rho$ in Fig. \ref{fig:model_2_cycles}b$_1$ and $p$ in Fig. \ref{fig:model_2_cycles}c$_1$). 
In Fig. \ref{fig:model_2_cycles} we compare the temporal evolution of jet solution in different cycles. At the end of first cycle the PSD returns to its original position ($\xs^\prime=25$) but the jet solution (plotted with blue color) is completely different in comparison to the steady state solution (magenta color) which was also obtained for the accretion disc when it was in steady state and $\xs=25$ (Figs. \ref{fig:model_2_cycles}a$_1$, b$_1$ and c$_1$). The time dependent jet harbours a time dependent shock at $r_{\rm js}=30$ at the end of the first cycle. At large distance the time dependent jet solution merges
with the steady state jet, since within the first cycle the information of an oscillating $\xs$ has not reached the entire length of the jet.
Figures \ref{fig:model_2_cycles} a$_4$, b$_4$ and c$_4$ compare the $v$, $\rho$ and $p$ distribution of the jet for the PSD size of $\xs^\prime=5.98$ in first and second cycle. It means $\xs^\prime$ is at its minima but at two different cycles of oscillation. The jet in the first and second cycles harbours a shock at $r_{\rm js}\sim 14$, while the jet in the second cycle harbours an additional shock at $r_{\rm js} \sim 400$. We again compare the jet in the equilibrium position of the accretion disc (i. e., when $\xs^\prime=25$)
but in the first and second cycle of oscillations (Fig. \ref{fig:model_2_cycles}a$_2$, b$_2$ and c$_2$), then again the jet
in the first cycle harbours one shock, while the one in the second cycle
has an additional shock. The first shock coincides for jets in both the cycles, but the second shock is present in the jet in the second cycle only. In the fifth column (Fig. \ref{fig:model_2_cycles}a$_5$, b$_5$ and c$_5$) we compare the jet in the second and third cycle but when $\xs^\prime=5.98$. The jet solution coincides. Similarly, when the PSD is in its equilibrium position, we compare the jets in the second and the third cycle and the jet solutions coincide, although the two shocks advance
in the forward direction (Fig. \ref{fig:model_2_cycles}a$_3$, b$_3$ and c$_3$).

In Fig. (\ref{fig:jetsok}a, b \& c) we plot the variation of the compression ratio $R$ of the jet shock, the jet shock location $r_{\rm js}$ and $\xs^\prime$
as a function of time for Model 2. The compression ratio is
given by $R=\rho_+/\rho_-$. In (\ref{fig:jetsok} c) we have also marked the location of $\xs^\prime$ corresponding to the solutions plotted in Fig. \ref{fig:model_2_cycles}, using the diamond markers, the color coding and labels of these markers are kept similar to the colors used to show the solutions in Figure (\ref{fig:model_2_cycles} $\rm{a_1-a_5}$), and $\rm{a_s}$ shows the location of $\xs$ for the steady state outflow solution (magenta, dotted) plotted in Fig. (\ref{fig:model_2_cycles} $\rm{a_1}$, b$_1$ \& c$_1$). The PSD luminosity remains same throughout the
oscillation. Although the accretion rate of the SKD remains same
but the oscillation of $\xs^\prime$ increases and decreases the SKD brightness with time. In the steady state regime, there was no shock in the jet. During the inbound path of $\xs^\prime$, the intensity of radiation increases and forms a shock in
the jet $r_{\rm js}$. The jet shock strength has complicated variation with time, however, tends to reach a asymptotic value as it leaves the computational domain. As the jet-shock from the first cycle moves out, a second shock develops at around the same phase of $\xs^\prime$ oscillation in the second cycle. As $\xs^\prime$ continues to oscillate one can witness multiple shocks forming in the jet and all of those shocks are moving outward. So disc oscillation would create time dependent radiation field and that can produce multiple jet shocks.
The shocks are moderately strong and it seems that $R>1.5$ as the shocks leave the computational domain at $r=1000 \rg$.

\section{Effect of composition parameter on jet solutions}

\begin{figure}
\includegraphics[width=\columnwidth]{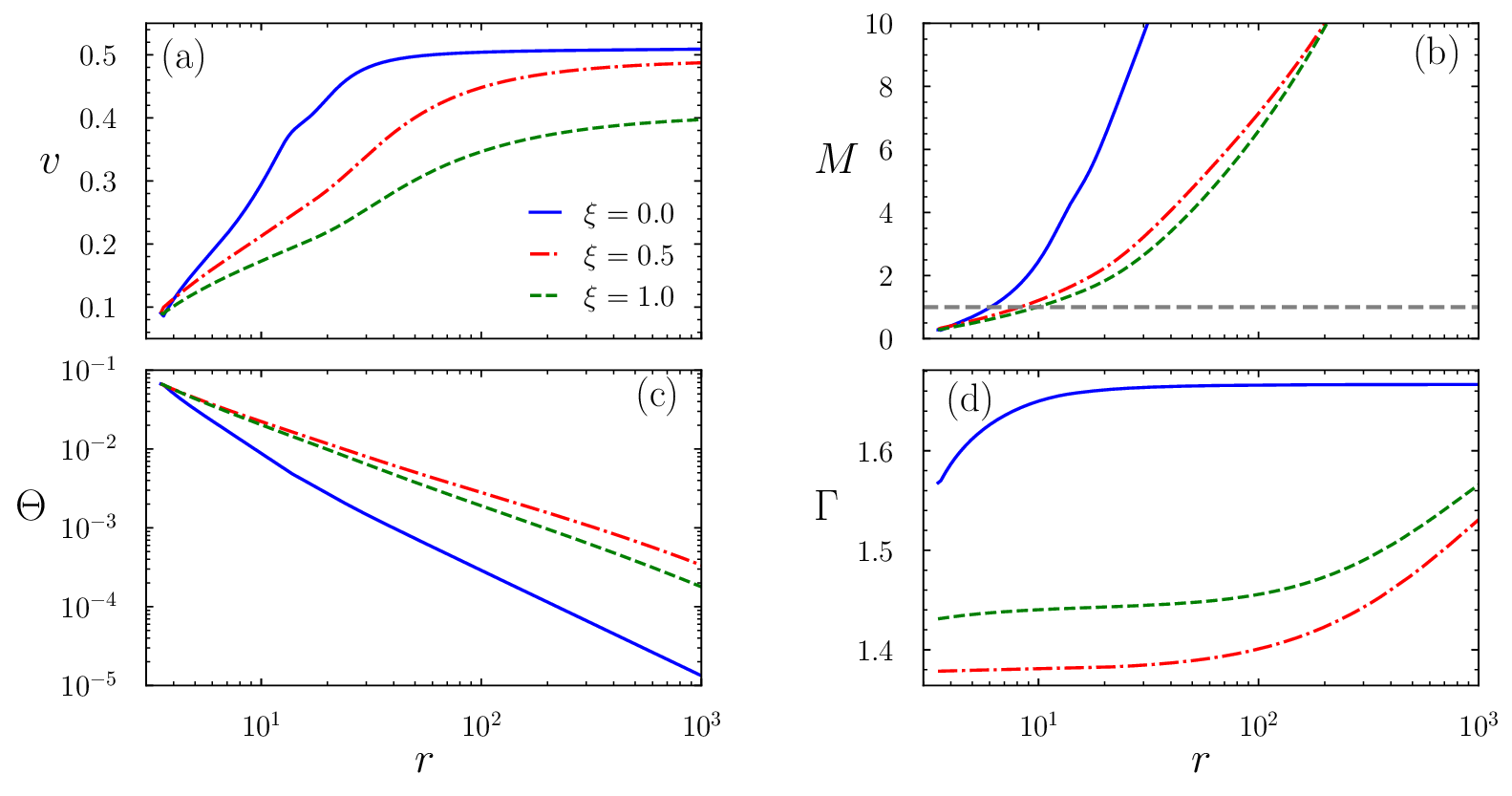}
\caption{Effect of composition parameter $\xi$ on jet solutions. The injection parameters are kept same for all three cases and composition parameter for each solution is mentioned in the legend.}
\label{fig:xi_effect_steady}
\end{figure}
The CR EoS used in this work allows us to study the effect of plasma composition ($\xi$) on the jet dynamics. To study the effect of composition we generate the solutions for different $\xi$ while keeping the disc and injection parameters same.

\begin{figure*}
\includegraphics[width=15cm]{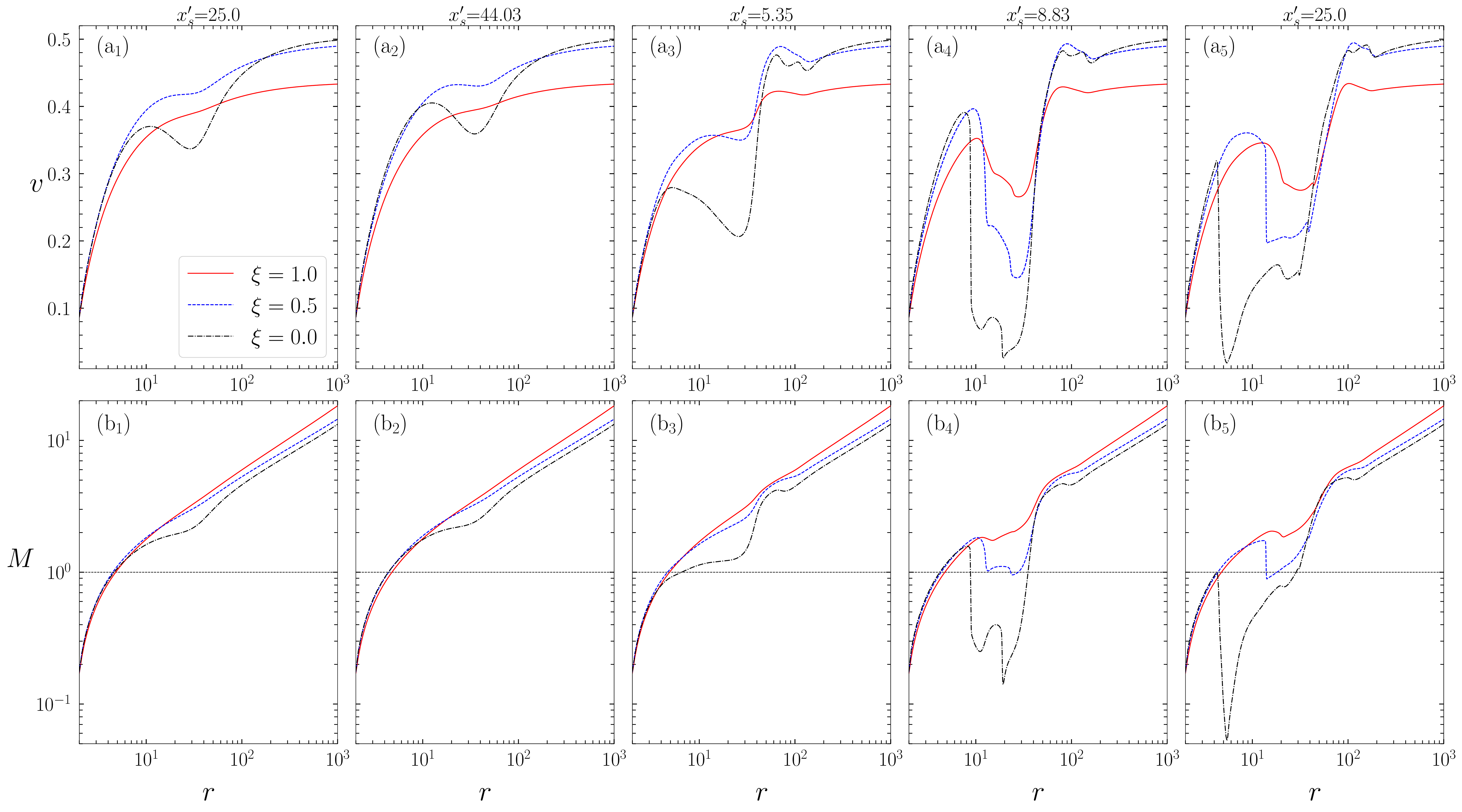}
\caption{Comparison of time dependent jet of the type Model 2: The PSD oscillates with the same amplitude of 20 and the time period $T=3.3\times 10^2$t$_g$.}
\label{fig:xi_effect_mod2}
\end{figure*}

In Fig. \ref{fig:xi_effect_steady} we have plotted the velocity ($v$), Mach number ($M$), temperature ($\Theta$), and adiabatic index ($\Gamma$) profiles for jet with $\xi=0.0$ (solid blue), $\xi=0.5$ (dash-dotted red), and $\xi=1.0$ (dashed green). The injection parameters $\vin=0.0906,\, \thin= 0.0669$ at $\rin=3.5$ were kept same for all three cases and disc parameters are similar to as given in equation \ref{eq:disc_par_thin}.
In other words, the disc parameters do not evolve in time. Moreover, the injection parameters of the flow are taken from analytical steady state solutions. Therefore it is expected that the steady state jet solutions would be regenerated by the simulation code. Figure \ref{fig:xi_effect_steady} represents the steady state solutions of
 jets with different composition. {$\xi=0$ corresponds to electron-positron jet}. The jet with lower value of $\xi$ for the same mass density will have the higher number of leptons in comparison to the jets with higher value of $\xi$. Because of the higher number of leptons present in the jet beam the momentum transferred to jet plasma increases resulting in higher velocity so the jet with $\xi=0$ is fastest. Also, the fluids with higher number of proton fraction will have less number of electrons and therefore the net momentum transferred from the radiation field to the jet will be less,
 so the jet with higher $\xi$ is slower and less hotter. 
 In Fig. \ref{fig:xi_effect_mod2} we compare the flow velocity $v$ (panels
 a$_1$---a$_5$) and Mach number $M$ (panels b$_1$---b$_5$)
 of jets with composition $\xi =1$ (red, solid), $\xi=0.5$ (blue, dashed) {and $\xi=0.0$
 (dash-dotted, black) with injection parameters $\vin=0.0867$, $\thin=0.18$ at $\rin=2$
 and disc parameters $l_{\rm ps}=0.4$ and ${\dot m}_{\rm sk}=5$.}
 The time period of oscillation of $\xs$ is $T=3.3\times 10^2 t_{\rm g}$.
 {The shock oscillation is inducing a shock transition in the jet, that effect is more pronounced for pair plasma.}
 Needless to say that the solutions depend significantly on their composition.
 
\section{Discussion and concluding remarks}
In this paper, we have studied the jets with spherical cross section under the influence of radiation supplied by the accretion disc. The radiation driven winds/jets has been studied extensively by Fukue
and his collaborators although the radiation field considered
by them, was generally from Keplerian discs and rarely from sub-Keplerian discs. We on the other hand tried to estimate the effect of radiation field produced by an advective and sub-Keplerian flow (the kind associated with low hard to hard intermediate states of micro-quasars). As the jet is fully ionized, the radiation field transfers momentum to the jet material through scattering. The thermodynamics of the outflow is described by an EoS with variable adiabatic index. The bipolar outflow is driven by the radiation of the accretion disc.
The disc plays an auxiliary role as the source of the radiation field and is not dynamically included in
the computation. To compute the radiation field, the 
accretion rate of the sub-Keplerian disc and the luminosity of the post-shock disc are supplied as free parameters. It may be noted that, the PSD luminosity
can be computed, but we have avoided such complications in the analysis. The chosen parameters are reasonable and agree with previous studies.
In this paper, we focused mainly on time dependent studies of such outflows. We have first generated semi-analytical steady state, radiatively driven fluid jets (bipolar outflows) and then chose
a launching radius and jet flow variables at that location as
injection parameters for the time dependent code. We have developed the time dependent code following standard TVD scheme, but also
used the CR EoS to describe the thermodynamics of the jet flow. The hydrodynamic test of the basic code is presented in the appendix as shock-tube test. But we have used the time dependent code in spherical coordinates to study radiatively driven jets. So to test the code in the spherical coordinate version, steady state jet solutions were used to test how well the time dependent code regenerates the analytical jets. We found that we need higher resolution to regenerate the jet.
In this paper, each cell corresponds to $\Delta x_i = 0.1666~\rg$,
the time resolution is obtained by CFL condition.

While matching the steady solution with the time dependent code as it reaches the steady state, we have shown that the radiation field plays an important role in the acceleration of the jet. We have also shown that inside the funnel region of PSD increment in the intensity of radiation field can accelerate or decelerate the jet material, depending upon the geometry of the disc. The radiation fields arising out of different components of the disc peak at different location of the jet axis, hence it can be responsible for the multistage acceleration of the jet. All these general results have been established via steady state jet solutions obtained earlier \citep{cc02b,vc18,vc19}. 

In this paper, the main focus is on how a time dependent radiation field may affect the jet. The time dependence of the radiation field has been induced through the oscillation of the disc, and not by time dependence of accretion rates. Further, in case the disc is oscillating then various parts of the jet will receive this information of the motion
of various parts of the disc, at different time. This happens because the radiation field travels with a finite speed. Although there are predictions of
radiatively driven stationary jet shocks \citep{vc18,vc19}, we simulated jets with injection parameters which would produce smooth
steady jets. If the PSD while oscillating remains geometrically thick then, the resulting radiation field will oppose the forward acceleration of the jet. In such cases jet shock might develop only at certain phases. However, if the accretion disc oscillates in a manner that the PSD becomes geometrically thick as the $\xs^\prime$ decreases,
then the radiation field drives a shock in the jet. But such jet-shock drifts outward as $\xs^\prime$ moves always, only to produce another jet-shock in the next cycle. In this way one can produce a large number of traveling shocks in the jet. So variable radiation field gives rise to jet solutions with multiple sonic points and time dependent shocks closer to the jet base. These internal shocks closer to jet base have been used to explain the high-energy power-law emission in the microquasars. \citep{lrwcpg11}. We also showed that jet solutions
differ significantly based on the composition of the flow. The lepton dominated jets are faster than electron-proton jets. Moreover, we also showed that time dependent radiation field can produce significantly different jet solutions.
\label{sec:disc}

\section*{Data Availability}
The data underlying this article will be shared on reasonable request to the corresponding author.



\bibliographystyle{mnras}
\bibliography{biblio} 




\appendix

\section{Radiative moments from SKD} \label{sec:app2}
The functions which mimic the radiative moments from SKD are given as 
\begin{equation}
\mathcal{E}=\mathcal{S}_{sk}\mathcal{C}_1(h_s)\left[E_f(r, x_o)-E_f(r, x_i)\right]    
\label{eq:eng_alb}
\end{equation}

Where
\begin{equation}
E_f(r, x)=-\frac{E_1(r,x)+E_2(r,x)}{E_3(r, x)}    
\end{equation}

\begin{eqnarray*}
E_1(r,x)=rA(r,x)^{1/2}{\rm{sinh}}^{-1}\left(\frac{r^2-xr\cot\thsk}{|x||r|}\right) \nonumber \\
E_2(r,x)=\left[r\left(\cot^2\thsk-1\right)-\left(\cot^3\thsk+\cot\thsk\right)x\right]|r| \nonumber \\
E_3(r,x)=|r|A(r,x)^{1/2}r^2 \nonumber \\
\end{eqnarray*}

\begin{equation}
\mathcal{F}=\mathcal{S}_{sk}\mathcal{C}_2(h_s)\left[F_f(r, x_o)-F_f(r, x_i)\right]    
\label{eq:forc_alb}
\end{equation}

Where
\begin{equation}
F_f(r, x)=-r\left[(F_1(r, x)+F_2(r,x)+F_3(r, x)+F_4(r,x)\right]    
\end{equation}

\begin{eqnarray*}
F_1(r,x)=-\frac{{\rm{log}}[A(r,x)]}{2r^3}  \nonumber \\
F_2(r,x)=\frac{1}{r^3}\cot\thsk\,{\rm{tan}}^{-1}\left[\frac{(1+\cot^2\thsk)x-r\cot\thsk}{r}\right] \nonumber \\ 
F_3(r,x)=\frac{1}{2rA(r,x)}   \nonumber \\  
F_4(r,x)=\frac{\rm{log}(x)}{r^3}  \nonumber \\ 
\end{eqnarray*}

\begin{equation}
\mathcal{P}=\mathcal{S}_{sk}\mathcal{C}_3(h_s)\left[P_f(r, x_o)-P_f(r, x_i)\right]    
\label{eq:pres_alb}
\end{equation}

Function $P_f(r, x)$ is given as 
\begin{equation}
P_f(r, x)=-\frac{P_1(r,x)+P_2(r,x)}{P_3(r, x)}    
\end{equation}

\begin{eqnarray*}
P_1(r,r)=3zA(r,x)^{3/2} {\rm{sinh}}^{-1}\left[\frac{r^2-xr\cot\thsk}{xr}\right] \nonumber \\
P_2(r,x)=|r|[3x^2r(\cot^3\thsk-1)-(\cot^5\thsk+2\cot^3\thsk+\cot\thsk)x^3  \nonumber \\
+r^3(\cot^2\thsk-4)+(6\cot\thsk-3\cot^3\thsk)xr^2] \nonumber \\
P_3(r,x)=3r^2A(r,x)^{3/2}|r| \nonumber
\end{eqnarray*}

Function $A(r,x)$ is given as
\begin{equation}
A(r, x)=\left(r-x\,\cot\thsk\right)^2+x^2 \nonumber   
\end{equation}
And factors $\mathcal{C}_i(h_s)$ (i=1, 2, 3) are given as  
\begin{equation}
C_i(h_s)=a_i{\rm exp}(-hs/b_i)+c_i/h_s^{3.5}    
\end{equation}
For the model when we consider a constant shock height the functions given in \ref{eq:eng_alb}, \ref{eq:forc_alb}, \ref{eq:pres_alb} still mimic the radiative moments with the only difference that the factors $\mathcal{C}_i$ depend upon the shock location $\xs$ instead of shock height, given as 
\begin{equation}
C_i(x_s)=a_i{\rm log}(x_s)+b_i    
\end{equation}
Values of $a_i, \, b_i, \, c_i$ are given below in table
\begin{tabular}{ |l||l|l|l|  }
 \hline
 \multicolumn{4}{|c|}{Model} \\
 \hline
Parameters & $h_s=r_s$ & $h_s=4r_s$ & $h_s=40$\\
 \hline
$a_1$  &$-7.02\times10^5$ & $8.73\times10^2$    &   $-4.49\times10^2$\\
$a_2$  &$-6.69\times10^5$ & $8.14\times10^2$    &  $-4.73\times10^2$\\
$a_3$  &$-6.84\times10^5$ & $8.36\times10^2$    &  $-5.07\times10^2$\\
$b_1$  &$2.52$            & $3.25\times10^1$    &   $2.29\times10^3$\\
$b_2$  &$2.67$            & $3.40\times10^1$    &  $2.45\times10^3$\\
$b_3$  &$2.86$            & $3.39\times10^1$    &  $2.60\times10^3$\\
$c_1$  &$1.90\times10^8$  & $4.07\times10^8$    &   -\\
$c_2$  &$2.14\times10^8$  & $4.48\times10^8$    &   -\\
$c_3$  &$2.39\times10^8$  & $4.74\times10^8$    &   -\\
 \hline
\end{tabular}


\bsp	
\label{lastpage}
\end{document}